\documentclass[prd,aps,twocolumn,nofootinbib,showpacs,amsmath,superscriptaddress,floatfix]{revtex4-1}

\usepackage{graphicx}
\usepackage{bm}
\usepackage{xspace}
\usepackage{amssymb}
\usepackage{amsmath}
\usepackage{upgreek}
\usepackage[dvipsnames]{xcolor}
\usepackage[utf8]{inputenc}
\usepackage[normalem]{ulem}
\usepackage{stmaryrd}
\usepackage[hidelinks]{hyperref}
\usepackage{multirow}

\newcommand{\mbar}{\overline{m}}

\def\OMIT#1{}

\newcommand{\df}{\mathrm{d}}

\newcommand{\nn}{\nonumber}

\newcommand{\bn}{{\bar n}}
\newcommand{\bea}{\begin{eqnarray}}
\newcommand{\eea}{\end{eqnarray}}
\newcommand{\tb}{{\bar{t}}}

\newcommand{\gsim}{\mathrel{\rlap{\lower4pt\hbox{\hskip1pt$\sim$}}\raise1pt\hbox{$>$}}}

\newcommand{\PythiaOld}{{\scshape Pythia}~8.205\xspace}

\newcommand{\shattau}{\hat{s}_\tau}

\newcommand{\vel}{\varrho}
\newcommand{\MS}{\overline{\rm MS}}

\DeclareRobustCommand{\Ref}[1]{Ref.~\cite{#1}}
\DeclareRobustCommand{\Refs}[1]{Refs.~\cite{#1}}

\DeclareRobustCommand{\eq}[1]{Eq.~\eqref{eq:#1}}
\DeclareRobustCommand{\eqs}[2]{Eqs.~\eqref{eq:#1} and \eqref{eq:#2}}
\DeclareRobustCommand{\fig}[1]{Fig.~\ref{fig:#1}}

\DeclareRobustCommand{\sec}[1]{Sec.~\ref{sec:#1}}

\newcommand{\app}[1]{App.~\ref{app:#1}}
\newcommand{\Mathematica}{\textsc{Mathematica}\xspace}
\newcommand{\PLHot}{\textsc{PLHot}\xspace}

\begin{document}

\title{Boosted Top Quarks in the Peak Region with N$^3$LL Resummation}
\author{Brad Bachu}
\affiliation{Center for Theoretical Physics, Massachusetts Institute of Technology, Cambridge, MA 02139, USA}
\affiliation{Department of Physics, Princeton University, Princeton, New Jersey, 08544, USA}
\author{Andr\'e H. Hoang}
\affiliation{University of Vienna, Faculty of Physics, Boltzmanngasse 5, A-1090 Wien, Austria}
\affiliation{Erwin Schr\"odinger International Institute for Mathematical Physics,\\
University of Vienna, Boltzmanngasse 9, A-1090 Wien, Austria}
\author{Vicent Mateu}
\affiliation{Departamento de F\'isica Fundamental e IUFFyM,\\Universidad de Salamanca, E-37008 Salamanca, Spain}
\affiliation{Instituto de F\'isica Te\'orica UAM-CSIC, E-28049 Madrid, Spain}

\author{Aditya Pathak}
\affiliation{Center for Theoretical Physics, Massachusetts Institute of Technology, Cambridge, MA 02139, USA}
\affiliation{University of Vienna, Faculty of Physics, Boltzmanngasse 5, A-1090 Wien, Austria}
\affiliation{University of Manchester, School of Physics and Astronomy, Manchester, M13 9PL, United Kingdom}
\author{Iain W. Stewart}
\affiliation{Center for Theoretical Physics, Massachusetts Institute of Technology, Cambridge, MA 02139, USA}

\begin{abstract}
We present results for the 2-jettiness differential distribution for boosted top quark pairs produced in $e^+e^-$ collisions in the peak region accounting for QCD large-logarithm resummation at next-to-next-to-next-to-leading logarithmic (N$^3$LL) order and fixed-order corrections to matrix elements at next-to-next-to-leading order (NNLO) calculated in the framework of soft-collinear effective theory and boosted heavy quark effective theory. Electroweak and finite-width effects are included at leading order. We study the perturbative convergence of the cross section in the pole and MSR mass schemes, with and without soft gap subtractions. We find that there is a partial cancellation between the pole mass and soft function renormalons. When renormalon subtractions concerning the top mass and the soft function are implemented, the perturbative uncertainties are, however, systematically smaller and an improvement in the stability of the peak position is observed. We find that the top MSR mass may be determined with perturbative uncertainties well below $100$\,MeV from the peak position of the 2-jettiness distribution. This result has important applications for Monte Carlo top quark mass calibrations.
\end{abstract}

\pacs{12.38.Bx, 12.38.Cy, 12.39.St, 24.85.+p
\hfill Preprint: MIT-CTP 5237, UWThPh-2020-24, IFT-UAM/CSIC-20-134, MAN/HEP/2020/012}

\maketitle

\section{Introduction}
The top quark mass $m_t$ is one of the most important parameters of the Standard Model (SM). In conjunction with the Higgs boson mass, it is an essential input for
studies of the stability of the SM electroweak vacuum~\cite{Cabibbo:1979ay,Alekhin:2012py,Buttazzo:2013uya,Andreassen:2014gha,Branchina:2013jra,Branchina:2014usa}, and it plays an important role in precision electroweak fits~\cite{Baak:2014ora}. The most precise determinations of the top mass to-date come from so-called ``direct measurements'', that are based on the kinematic reconstruction of the final-state top quark decay products and the comparison of the resulting kinematic distributions with parton-shower Monte Carlo (MC) simulations. The current world average for direct measurements reads \mbox{$m_t^{\rm MC} = 172.76 \pm 0.30$\,GeV}~\cite{Zyla:2020zbs} and projections for the \mbox{HL-LHC} indicate that uncertainties as small as $200$\,MeV can be reached for individual measurements~\cite{Azzi:2019yne}.

The interpretation of these measurements are, however, (as reviewed below) impacted by an additional ambiguity from the lack of understanding of the field theoretic meaning of the top mass parameter encoded in the MC event generators~\cite{Hoang:2008xm,Hoang:2014oea,Corcella:2019tgt,Azzi:2019yne}. This ambiguity is not yet precisely quantified and should be considered at the GeV level, i.e.\ it is comparable to the uncertainties quoted by the experimental analyses~\cite{Hoang:2020iah}. Carrying out first-principle theoretical predictions of kinematic distributions that exhibit high sensitivity to the top mass is a challenging program in the light of the disparate energy scales that enter the top production and the measurements on the decay products. These lead to large logarithms of ratios of these scales that require resummation. Furthermore, due to nonperturbative corrections it is necessary to take into account the hadronic nature of the final state. While these effects can be accounted for via MC event generators, it comes at the cost of the limited perturbative~\cite{Dasgupta:2018nvj,Hoang:2018zrp} and conceptual precision of the MC description of perturbative and nonperturbative effects. This is the origin of the interpretation problem of the top mass parameter employed in the MC generators (see~\cite{Hoang:2020iah} for further discussion).

In order to achieve a precise top quark mass determination we therefore need an observable that has the required kinematic sensitivity and is theoretically tractable, such that (a)~it can be reliably calculated in perturbation theory within a specific short-distance top quark mass scheme, and (b)~nonperturbative effects can be consistently quantified from a field theory perspective. Such calculations can be carried out in the framework of effective field theories (EFTs) that are systematically improvable in their power counting expansion, as well as in perturbation theory where resummation of large logarithms now reaches next-to-next-to-next-to leading logarithmic (N$^3$LL) accuracy for a number of applications~\cite{Becher:2008cf,Chien:2010kc,Abbate:2012jh,Hoang:2014wka}. Moreover, the framework of EFTs offers ways to rigorously describe and quantify nonperturbative effects due to hadronization~\cite{Lee:2006fn,Hoang:2007vb,Mateu:2012nk,Hoang:2019ceu}.

Therefore, for such an observable, EFT based calculations may naturally incorporate all the distinct features of parton-shower MC simulations while retaining a systematic connection to field theory and thus theoretical control. Hence, EFTs offer promising prospects for precision collider physics and for developing diagnostic tools for improving MC simulations~\cite{Mo:2017gzp,Hoang:2019ceu}. Such a framework has recently been applied to the broad program of top quark mass measurements, including a factorized description of a hadron-level differential top jet mass spectrum for jets initiated by boosted top quarks in the peak region at a future $e^+e^-$ collider~\cite{Fleming:2007qr,Fleming:2007xt}, boosted top jets with soft drop grooming at the LHC~\cite{Hoang:2018zrp}, a calibration of the MC top quark mass parameter~\cite{Butenschoen:2016lpz} (based on the work of Refs.~\cite{Fleming:2007qr,Fleming:2007xt}), and parton-level studies of the correlation of the MC top quark mass parameter with the parton shower evolution cutoff~\cite{Hoang:2018zrp}. The sensitivity of event-shape variable definitions with respect to quark mass effects has also been recently studied in Refs.~\cite{Lepenik:2019jjk,Bris:2020uyb} in the context of fixed-order and resummed perturbation theory.

In this work we continue this effort by extending the perturbative calculations of boosted tops in the peak region at an $e^+e^-$ collider from the N$^2$LL resummation of logarithms and ${\cal O}(\alpha_s)$ fixed-order matrix elements used in~\cite{Butenschoen:2016lpz}, to N$^3$LL resummation with NNLO [\,${\cal O}(\alpha_s^2)$\,] fixed-order matrix elements.

\subsection{Status of top mass measurements}

Recent direct top quark mass measurements have yielded the results $m_t^{\rm MC} = 172.26 \pm 0.61$\,GeV (CMS)~\cite{Sirunyan:2018mlv}, $m_t^{\rm MC} = 172.69 \pm 0.48$\,GeV (ATLAS)~\cite{Aaboud:2018zbu} at the LHC, and \mbox{$m_t^{\rm MC} = 174.34 \pm 0.64$\,GeV}~\cite{Tevatron:2014cka} at the Tevatron. The top mass superscript MC signifies that the direct measurements extract the top quark mass parameter coded in the MC generators used for the analyses. The lack of understanding of a precise field-theoretic definition of $m^{\rm MC}_t$ (and hence its relation to short-distance masses defined in the context of quantum filed theory, which is the preferred input parameter for high-precision theoretical predictions) results in an additional conceptual uncertainty in how $m^{\rm MC}_t$ should be related to the pole mass $m_t^{\rm pole}$ or a short-distance mass such as the $\overline{\rm MS}$ mass $\overline m_t^{(6)}(\mu)$ or the MSR mass $m_t^{\rm MSR}(R)$~\cite{Hoang:2008yj,Hoang:2017suc,Hoang:2017btd}. This ambiguity is not included in the individual quoted experimental uncertainties but should be considered to be (at least) comparable~\cite{Hoang:2020iah}. Analyses that can shed light on a more complete and quantitative understanding of these issues, as well as more first-principle aspects of MC event generators, are underway~\cite{Hoang:2018zrp,Hoang:2017kmk,Butenschoen:2016lpz}. Other recent related studies include analyses of the theoretical limitations concerning the modeling of the dynamics in the top quark production and decay~\cite{FerrarioRavasio:2019vmq}, finite lifetime~\cite{Heinrich:2017bqp}, hadronization effects~\cite{Corcella:2017rpt} and observable infrared sensitivity~\cite{Nason:2019gff}.

The so-called top quark pole-mass measurements based on the total cross section~\cite{Czakon:2013goa}, for which precise theoretical predictions expressed in terms of the pole mass renormalization scheme have been employed, have generally yielded results with larger uncertainties than the direct measurements: \mbox{$m^{\rm pole}_t= 172.9^{+2.5}_{-2.6}$\,GeV} (ATLAS)~\cite{Aad:2014kva} and \mbox{$m^{\rm pole}_t= 172.7^{+2.4}_{-2.7}$\,GeV} (CMS)~\cite{Sirunyan:2017uhy}.
The relatively large errors result from the uncertainty in the normalization of the inclusive cross section (both in the measurement and the theoretical prediction) and its relatively weak dependence on $m_t$. Employing differential cross sections such as leptonic distributions, and using matched NLO+PS (parton shower) MC simulations results in an enhanced top mass sensitivity~\cite{Aaboud:2017ujq}. Such differential measurements have been included in the world average \mbox{$m_t^{\rm pole} = 172.4 \pm 0.7$\,GeV}~\cite{Zyla:2020zbs} for the pole mass measurements which is in good agreement with the corresponding world average of direct measurements mentioned above. Recent precision measurements, also using matched NLO+PS MC simulations for the theoretical predictions, accounting for $t\bar t+$jet final states~\cite{Alioli:2013mxa} ($m^{\rm pole}_t= 171.1^{+1.2}_{-1.0}$\,GeV~\cite{Aad:2019mkw}) and the reconstructed top-antitop invariant mass distribution (\mbox{$m^{\rm pole}_t= 170.5\pm 0.8$\,GeV} from a simultaneous $\alpha_s$ fit~\cite{Sirunyan:2019zvx}) have comparable uncertainties but have, however, resulted in significantly lower $m_t$ values, posing some tension that may partly arise from missing theoretical input in predictions for the corresponding differential cross sections~\cite{Hoang:2020iah,Ju:2020otc,Catani:2020tko}.

It is well known that the pole mass additionally suffers from a conceptual ${\cal O} (\Lambda_{\rm QCD})$ renormalon ambiguity which in dedicated analyses was estimated to amount to $110$\,MeV in Ref.~\cite{Beneke:2016cbu} and to $250$\,MeV in Ref.~\cite{Hoang:2017btd}.\footnote{The quoted numbers for the pole mass renormalon ambiguity arise in the context of finite bottom and charm masses. When charm and bottom quarks are treated as massless quarks, the infrared sensitivity decreases and pole mass ambiguity is smaller.} This ambiguity does not represent an uncertainty due to perturbative truncation or limited information, but signifies the principle conceptual imprecision in assigning a definite value to $m^{\rm pole}_t$. The ambiguity is not related to any physical effect, but inherent to the unphysical nature of the pole mass renormalization condition. It can therefore be avoided by expressing cross sections obtained in perturbation theory in terms of a short-distance mass scheme at an appropriate renormalization scale, which can also improve the overall convergence of the perturbative series at the first few orders. As far as the spread of the above-mentioned recent pole mass measurements is concerned, the pole mass renormalon ambiguity likely plays no role, because it is smaller than the quoted uncertainties of these measurements. For the direct top quark mass measurements, the pole mass renormalon problem has been discussed intensely in the context of the frequently used approach of identifying $m_t^{\rm MC}$ and $m_t^{\rm pole}$. However, in Ref.~\cite{Hoang:2018zrp} it was shown analytically for the simple case of the 2-jettiness distribution in $e^+e^-\to t\bar t+X$, that the quark mass parameter associated to a NLL-precise parton shower is in general not the pole mass, but a low-scale short distance mass that depends on the value of the shower cut and may differ from the pole mass by an amount larger than the pole mass ambiguity.\footnote{In Ref.~\cite{Hoang:2018zrp} it was shown that the quark mass parameter associated to a NLL-precise parton shower based on the coherent branching algorithm with shower cut $Q_0$, is a \mbox{$Q_0$-dependent} short-distance mass $m_t^{\rm CB}(Q_0)$ that differs from the pole mass by the amount \mbox{$m_t^{\rm CB}(Q_0)-m_t^{\rm pole}=-(2/3)\alpha_s(Q_0) Q_0$}.} So the problem of how to properly interpret the MC top quark mass $m_t^{\rm MC}$ in terms of a well-defined and ambiguity-free field theory mass is not related to the pole mass renormalon ambiguity, but to the limited theoretical precision of the state-of-the-art MC event generators and to ignorance concerning MC systematics.

Motivated by the interpretation issues of the direct top mass measurements and the still large uncertainties in the pole mass measurements from inclusive and differential cross sections, a number of alternative methods to measure $m_t$ have been proposed, which are based on differential cross sections with respect to variables constructed from top decay products exhibiting strong kinematic top mass sensitivity. The observables these analyses are based on include the $M_{T2}$ variable and variants of it~\cite{Lester:1999tx,Chatrchyan:2013boa}, the lepton $b$-jet invariant mass~\cite{CMS:2014cza}, the shape of $b$-jet and $B$-meson energy distributions~\cite{Agashe:2016bok}, and the $J/\psi$ and lepton invariant masses~\cite{Khachatryan:2016pek,Khachatryan:2016wqo}. Conceptually, these observables are governed by parton shower dynamics as well as various nonperturbative effects, in a way analogous to the direct reconstruction method (albeit with differing systematics). This is because they are also based on the idea of assessing the kinematics of decaying (colored) top quark particles through simulations obtained from parton-shower MCs. Their reliance on these MCs further makes their potential extension to higher logarithmic precision nontrivial, as the theoretical precision of parton showers is quite observable-dependent. Additionally, the nonperturbative effects of hadronization are accounted for through a multi-parameter MC hadronization model. Here, since a systematic way to quantify the intrinsic MC hadronization uncertainty does not yet exist, uncertainties are typically being estimated by comparing predictions based on different models.

\subsection{Top mass determination using effective field theories}

In \Refs{Fleming:2007qr,Fleming:2007xt} a framework of EFTs was developed to describe boosted top quarks in the peak region at a future $e^+e^-$ collider. Using soft-collinear effective theory (SCET)~\cite{Bauer:2000ew,Bauer:2000yr,Bauer:2001yt,Bauer:2001ct,Bauer:2002nz} a factorization theorem for the double-differential hemisphere-mass cross section in $e^+e^-\rightarrow t\tb+X$ was derived in the boosted top quark limit, with center of mass energy $E_{\rm cm}=Q \gg m_t$. The two invariant masses $M_t$ and $M_\tb$ are defined using all particles in each of the two hemispheres that are determined by the event's thrust axis described below. The peak refers to the region where the $M_t$\,--\,$M_\tb$ double differential distribution exhibits the hemisphere mass top and antitop resonances. One key feature of this factorization formula is that the most important hadronization effects are parametrized via a convolution with a top-mass- and $Q$-independent nonperturbative shape function, which is field theoretically defined from a vacuum matrix element of Wilson-lines. The universality of the factorization formula states that, at least in principle, data for massless dijet events (obtained even at past experiments such at LEP) could be used to fix the nonperturbative shape function and to make the analysis independent of estimates of nonperturbative corrections obtained from MC event generators. The other key feature is that the dependence on the renormalization scheme of the top quark mass is fully controlled so that one can make and test predictions in any scheme to the extent that higher order perturbative corrections are incorporated. In the tail of the distribution an operator expansion can be applied such that the nonperturbative corrections are dominated by a single parameter.

The dijet hemisphere mass cross section exhibits a clear peak at the top and antitop resonances which are directly sensitive to the value of the top quark mass. However, the resonance location in $M_t$ and $M_\tb$ is not directly at the top quark mass due to radiative effects related to ultra-collinear and large-angle soft radiation, as well as hadronization and finite width effects. The ultra-collinear radiation refers to radiation that is soft in the (anti)top quark rest frame, but becomes collinear due to the (anti)top quark boost. In the factorization theorem this ultra-collinear radiation is described by boosted versions of heavy quark effective theory (HQET) \cite{Eichten:1989zv,Isgur:1989vq,Isgur:1989ed,Grinstein:1990mj,Georgi:1990um} called boosted HQET or simply bHQET.\footnote{Here the letter ``b'' in bHQET stands for the fact that the two HQET theories for the top and antitop quarks are boosted in opposite directions.} The typical scales of the ultra-collinear radiation in the peak region range between the top width $\Gamma_{\!t}$ and $\sim 10$\,GeV, which also quantifies the typical `off-shellness' $\sim(q^2-m_t^2)/m_t$ of the decaying top quarks. The top mass dependence of the location and shape of the observable peak, along with the top quark scheme dependence, is specified by ultra-collinear radiation effects which can be calculated perturbatively. Furthermore, the leading-order electroweak effects come from the top width which is fully encoded in Breit-Wigner-modified top quark propagators. The large-angle soft radiation is only sensitive to the collinear top quark color flow (which is fully taken over by the top decay products in the boosted limit) and describes soft momentum exchange between the two hemispheres. It is governed by scales below the top quark width, and its nonperturbative contributions constitute the shape function. Due to the boost of the top quarks, the effect of large-angle soft radiation on the peak locations is enhanced by a factor $Q/m_t$.

The above-mentioned framework was employed in \Ref{Butenschoen:2016lpz}, where a calibration of the $m_t^{\rm MC}$ parameter in \PythiaOld was carried out using the \mbox{N$^2$LL\,+\! $\mathcal{O}(\alpha_s)$} prediction for the 2-jettiness event shape in \mbox{$e^+e^- \rightarrow t\tb+X$} collisions defined as
\begin{equation}
\label{eq:tau2def}
\tau_2 \equiv 1- \max_{\hat n_t} \frac{\sum_i |\hat n_t \cdot \vec p_i|}{Q} \, ,
\end{equation}
where the sum runs over all produced particles $i$ in the event and $\hat n_t$ is the thrust axis that maximizes the sum in \eq{tau2def}. Since a sum over all final-state momenta is involved in \eq{tau2def}, we restrict ourselves to hadronically decaying top and anti-top jets, which is accomplished by simply including the corresponding branching fraction in the Born cross section $\sigma_0$ in the factorization formula discussed below. In the limit $\tau_2-2m_t^2/Q^2 \ll 1$, i.e.\ when focusing on the peak region, the event shape corresponds (up to power corrections) to the sum of the hemisphere invariant masses, such that
\begin{align} \label{eq:tauM}
\tau_2 = \frac{M_t^2+M_{\bar t}^2}{Q^2} + {\cal O}(\tau_2^2) \,,
\end{align}
and thus has the same kinematic sensitivity to the top mass as the double differential cross section considered in \Refs{Fleming:2007qr,Fleming:2007xt}. Being a global event shape, the $\tau_2$ differential cross section is furthermore free from non-global logarithms~\cite{Dasgupta:2001sh,Dasgupta:2002dc}. The results of the calibration carried out in \Ref{Butenschoen:2016lpz} showed that $m_t^{\rm MC}$ in \PythiaOld cannot be simply identified with the pole mass, but is numerically close to the MSR mass $m_t^{\rm MSR}(1\,\mbox{GeV})$. Compatible numerical results were obtained in the analysis of jets from top quarks produced in $pp$ collisions with soft drop grooming in \Ref{Hoang:2017kmk} based on theoretical calculations at NLL order.

In this work we improve the calculation of the 2-jettiness event shape cross section in $e^+e^- \rightarrow t\tb+X$ by including NNLO fixed-order matrix elements with N$^3$LL resummation [\,referred to as N$^3$LL\,+\,$\mathcal{O}(\alpha_s^2)$\,]. By now, all the ingredients that enter the factorization formula in \eq{fact} below are individually known to the accuracy needed to achieve this precision for the 2-jettiness cross section, in particular the two-loop heavy quark jet function~\cite{Jain:2008gb}, full two-loop thrust soft function~\cite{Monni:2011gb,Kelley:2011ng}, and the N$^3$LL result for the Wilson coefficient for the matching at the top mass scale~\cite{Hoang:2015vua}. We also include the recently calculated analytic 4-loop result for the cusp anomalous dimension~\cite{Henn:2019swt}, which despite having a tiny numerical impact, is an important formal ingredient to obtain N$^3$LL accuracy. We consistently combine these ingredients to obtain the N$^3$LL\,+\,$\mathcal{O}(\alpha_s^2)$ boosted top cross section fully analytically, and study the convergence of the resummed perturbation theory. We exclusively consider the $e^+e^-$ \mbox{2-jettiness} cross section in the boosted and bHQET limits where power corrections in $m_t/Q$ and in the top quarks' off-shellness $\sim(q^2-m_t^2)/m_t$ are neglected. These corrections, even though there are formally power-suppressed, can be non-negligible for phenomenological analyses in the peak region. With inclusion of these power corrections, the results obtained in this article will serve to improve analyses such as the top quark mass calibration carried out in \Ref{Butenschoen:2016lpz}, which we leave to future work.

The outline of the paper is as follows: We first present the factorization formula in \sec{fact} describing its key features and elements. We describe in \sec{mass} the implementation of the cross section in terms of short distance mass schemes and a renormalon-free soft function in \sec{gap}. The setup of observable-dependent renormalization scales is discussed in \sec{profile}. Finally, in \sec{numerics} we combine all the pieces to calculate the resummed cross section and study its perturbative convergence when the renormalons in the pole mass and the soft function are either subtracted or left unsubtracted. Using an analysis of the 2-jettiness peak location we draw conclusions on the perturbative uncertainties of a determination of the MSR mass and the pole mass. In the appendices we review and state additional details for the various ingredients that are needed in this analysis. We conclude in \sec{conclusion}.

\section{Factorization theorem in the peak region}
\label{sec:fact}
In Ref.~\cite{Fleming:2007qr} two factorized expressions for the differential cross section were derived that are valid in the peak and tail regions of the double hemisphere mass, or equivalently 2-jettiness, distribution. In the tail region the fluctuations in the top mass can be large, such that \mbox{$M_{t,\tb}^2 - m_t^2 \sim m_t^2$}, and a factorization formula based on SCET with massive particles can be derived. On the other hand, in the peak region, the off-shellness of the decaying top is constrained such that $M_{t,\tb}^2 - m_t^2 \ll m_t^2$, necessitating an additional factorization and a resummation accounting for the top width as an additional relevant scale that is carried out in the bHQET framework. As already mentioned, in this work we focus on the factorization in the peak region accounting exclusively for the bHQET contributions, leaving off-shellness power corrections described in the SCET factorization (that become essential in the tail region) and the inclusion of $m_t/Q$ SCET power corrections to future work. The factorization formula in the peak region is given by
\begin{align}\label{eq:fact}
&\frac{1}{\sigma_0}\frac{{\rm d} \sigma}{{\rm d}\tau_2} = m_t\, Q^2 H^{(5,6)}_{\rm evol}\! (Q, m_t, \vel, \mu ; \mu_H, \mu_m)
\\
& \!\times \!\! \int \! {\rm d}\ell\, {\rm d}\hat s\, U_B^{(5)}(\shattau\! - \!\vel \ell\! -\! \hat s, \mu, \mu_B)\,
J_{B,\tau_2}^{(5)}\!( \hat s , \Gamma_{\!t} , \delta m, \mu_B)
\nn \\
&\!\times\!\! \int \! {\rm d}\ell^\prime \df k\, U_S^{(5)}\!(\ell - \ell^\prime, \mu, \mu_S) \hat S_{\tau_2}^{(5)}(\ell^\prime - k , \bar \delta, \mu_S)
F(k - 2\Delta) \, ,\nn
\end{align}
where we have the boost parameter $\vel$ given by
\begin{align}\label{eq:velDef}
\vel \equiv \frac{Q}{m_t}\,,
\end{align}
and have defined the off-shellness variable $\hat s_\tau$ as\footnote{The limit $\hat s_\tau\to 0$ corresponds to the tree-level kinematics for $e^+e^-\to t\bar t$ where $\tau_2=2m_t^2/Q^2$.}
\begin{align}\label{eq:shatdef}
\shattau \equiv \frac{Q^2 \tau_2- 2 m_t^2}{m_t} \,.
\end{align}
Eq.~\eqref{eq:fact} involves various perturbative ingredients, including an evolved matching function $ H^{(5,6)}_{\rm evol}$, jet and soft functions $J_{B,\tau_2}^{(5)}$ and $\hat S_{\tau_2}^{(5)}$, and evolution kernels $U_B^{(5)}$ and $U_S^{(5)}$. It also includes a non-perturbative shape function $F$, whose independence from other parameters is a prediction of the factorization theorem. These ingredients will be discussed in detail in subsections below. Eq.~\eqref{eq:fact} applies in the peak region where $\hat s_\tau \sim \Gamma \ll m_t$, with $\Gamma \gtrsim 2\Gamma_{\!t}$ being the effective width of the distribution broadened by radiation, hadronization as well as finite width effects. In this region, where the $\tau_2$ distribution exhibits a resonance, the 2-jettiness variable is (up to power corrections) directly related to the sum of the squared hemisphere masses defined with respect to the thrust axis, as given in \eq{tauM}.
It is therefore convenient to define the inclusive jet mass variable $M_J$
\begin{align}
\label{eq:MJdef}
M^2_J \equiv \frac{1}{2} Q^2 \tau_2 \, ,
\end{align}
which inherits some of the features of a reconstructed top invariant mass, albeit being based on a hemisphere top jet. The $M_J$-distribution peaks close to $m_t$, but is in addition affected (with respect to peak position as well as the width of the observed peak resonance) by large-angle soft radiation exchanged between the two hemispheres. The widening of the peak due to top-decay width and soft QCD effects, however, does not affect the kinematic sensitivity of the observable, and normalizing the $M_J$ distribution enables the uncertainties in the $M_J$ peak location to be taken as a direct measure for the uncertainties in the associated top mass determination.

The factorization formula separates perturbative contributions from the hard local interactions involving the scales $Q$ and $m_t$ (encoded in the hard factor $H^{(5,6)}_{\rm evol}$), dynamical effects associated to large-angle soft radiation (accounted for in the soft function $S_{\tau_2}^{(5)}$), and dynamical effects due to ultra-collinear radiation (contained in the jet function $J_{B,\tau_2}^{(5)}$). The jet function $J_{B,\tau_2}^{(5)}$ incorporates the leading-order effects due to the top quark width $\Gamma_{\!t}$ and also carries, because of its peaked structure, the main top quark mass sensitivity of the $\tau_2$ distribution. This allows testing at high precision the impact of either using the pole mass scheme $m_t^{\rm pole}$ or a suitable short-distance mass $m_t^{\rm sd}$. This is indicated by the argument $\delta m_t$, where\footnote{In the context of HQET, the mass scheme correction $\delta m$ is called the residual mass term.}
\begin{align}\label{eq:deltam}
\delta m \equiv m_t^{\rm pole} - m_t^{\rm sd} \,,
\end{align}
is the perturbative series for the difference between the pole and the adopted short-distance masses.
The choice $\delta m=0$ implies the use of the pole mass scheme.
We note that there is also top quark mass (scheme) dependence in the hard factor $H^{(5,6)}_{\rm evol}$, indicated by the argument $m_t$, which, however, only affects the normalization of the $\tau_2$ distribution and is very subdominant compared to the main sensitivity to the top quark mass. The different character of the top mass dependence in the hard and the bHQET jet functions is discussed in more detail below.

For the $\tau_2$ distribution in the peak region for boosted top quarks we have the hierarchy $Q\gg m_t \gg \{\hat s_\tau,\Gamma\}$.
This leads to large logarithms, which are resummed via
renormalization group (RG) equations for the corresponding perturbative matrix elements.
This resummation is implemented through the evolution factors $U_i^{(n)}$ ($i=H,v$ (or $m),S,B$), which RG-evolve each of the functions appearing in the factorization theorem from their natural scales $\mu_i$ to a common final scale $\mu$. The quantities $\mu_i$ ($i=H,m,S,B$) are renormalization scales indicating the natural physical values of the quantum fluctuations encoded in the respective factors. These scales are varied in the final results in order to assess the theoretical uncertainties due to missing higher order contributions. The choice of $\mu$ is arbitrary and the factorized prediction is (strictly) invariant under changes of this $\mu$. In contrast the dependence on the initial scales $\mu_i$ only cancels out order-by-order in resummed perturbation theory. The scale $\mu$ is typically set equal to one of the renormalization scales $\mu_i$ ($i=H,m,S,J$) such that one of the renormalization evolution factors disappears.

The superscripts ``$(5)$'' and ``$(5,6)$'' indicate the number of active dynamical flavors relevant for the momentum scales of the respective quantum effects, where ``$(5)$'' and ``$(6)$'' indicate scales below and above the top mass, respectively.

Due to the simple inclusive character of the 2-jettiness (or the $M_J$) distribution, the leading-order nonperturbative effects arise from the low-energy dynamics of the large-angle soft radiation and are encoded in the (hadronization) shape function $F$, which is convolved with the perturbative soft function $S_{\tau_2}^{(5)}$. Physically, the shape function incorporates the leading effects of hadronization and controls the amount of nonperturbative radiation being exchanged between the two hemispheres.
Even though the details of the shape function form must be determined from experimental data, the way how the shape function appears in the factorized cross section represents a very strong theoretical constraint on hadronization.

In the following we briefly review the physical aspects of all functions appearing in the factorization theorem one by one. For a detailed discussion on how the bHQET factorization theorem of \eq{fact} connects to the corresponding formula in the tail region, and on possible alternative versions to organize the renormalization group evolution, we refer to Ref.~\cite{Hoang:2019fze}.

\subsection{The Hard Function}
The factorized cross section involves a two-step matching from QCD to SCET, and then from SCET to bHQET, at the scales $\mu_H \sim Q$ and $\mu_m \sim m_t$, respectively. Therefore, the difference between $6$ or $5$ flavors is related to the top quark being a dynamical degree of freedom or not. The resulting hard matching coefficients in \eq{fact}, together with their renormalization group (RG) evolution kernels, are collectively written as
\begin{align}
\label{eq:Hevol}
&H_{\rm evol}^{(5,6)} \!(Q, m_t, \vel, \mu ; \mu_H, \mu_m) \equiv H_Q^{(6)}(Q, \mu_H) \\
& \times U_{H_Q}^{(6)}(Q, \mu_H, \mu_m)\,H_m^{(6)}(m_t, \vel,\mu_m)\, U_v^{(5)} (\vel, \mu_m ,\mu ) \, .\nn
\end{align}
The term $U_{H_Q}$ evolves the SCET hard function $H_Q$~\cite{Matsuura:1987wt,Matsuura:1988sm,Gehrmann:2005pd,Moch:2005id,Baikov:2009bg,Lee:2010cga} from $\mu_H\sim Q$ to $\mu_m \sim m_t$ and resums large logarithms of $Q/m_t$ in the cross section. $U_v$ is responsible for the evolution of the bHQET current between $\mu_m \sim m_t$ and $\mu$, which we assume is smaller than $\mu_m$, and only depends on the top quark boost factor $\vel$ defined in Eq.~\eqref{eq:velDef}. The hard matching at the top quark mass scale $\mu_m$, which encodes off-shell top quark quantum fluctations that arise in the heavy quark limit, is given by $H_m$~\cite{Fleming:2007xt,Hoang:2015vua}. Since the matching is performed at the top quark mass, one can express $H_m$ in terms of $\alpha_s$ with either $5$ or $6$ active flavors. In the numerical analysis below we choose 6. The effect of this freedom in the scheme choice is, however, tiny and numerically irrelevant.

We now discuss the parameter $\vel=Q/m_t$ [\,see Eq.~\eqref{eq:velDef}\,] appearing in the mass mode matching factor $H_m^{(6)}$ and the bHQET current evolution kernel $U_v^{(5)}$. In the peak region, for the bHQET factorization treatment of the top quark dynamics, the momentum of the nearly on-shell top quarks is parameterized as $p_{t,\bar t}^\mu = m_t v_{t,\bar t}^\mu + k^\mu$, where $v_{t,\bar t}^\mu$ (with $v_{t,\bar t}^2 = 1$) is the $4$-velocity of the energetic (anti)top quarks
and $k^\mu$ is a small residual momentum accounting for the fluctuations caused by the low-energy radiation [\,in the (anti)top quark rest frame\,], such that $k^\mu \ll m_t$ and one can expand the dynamical effects to leading power in $k^\mu/m_t\sim \Gamma_{\!t}/m_t$ [\,in the (anti)top rest frame\,].
The reference $4$-velocities $v_{t,\bar t}^\mu$ are defined by
\begin{align}\label{eq:velttbar}
v_t^\mu & = (\vel^{-1}, \vel, \vec 0_\perp)\,,\\\nonumber
v_\tb^\mu & = (\vel, \vel^{-1}, \vec 0_\perp)\,,
\end{align}
using light-cone coordinates defined relative to the thrust axis $n^\mu = (1, \vec n_t)$ (which we take along the top direction), such that $p^\mu = (n \cdot p, \bn \cdot p, \vec p_\perp)$, where $\bn^\mu = (1, -\, \vec n_t)$ is an auxiliary vector satisfying $n \cdot \bn = 2$ and \mbox{$n^2=\bn^2=0$}. The parameter $\vel$ appearing in the definition of the reference velocities is related to the top quark boost and defined in Eq.~\eqref{eq:velDef},
such that the on-shell (anti)top $4$-velocity would approach $v_{t\bar t}^\mu$ in the boosted limit $Q\gg m_t$ in the absence of any radiation. The mass mode matching factor $H_m^{(6)}$ and the bHQET current evolution kernel $U_v^{(5)}$ depend on the reference velocities $v_{t}^\mu$ and $v_{\bar t}^\mu$ and thus on $\vel$ through the scalar products $v_{t,\bar t}\cdot k$ appearing in the bHQET Feynman diagrams.

The choice of $v_{t,\bar t}^\mu$ is ambiguous with respect to higher-order power corrections of ${\cal O}(k^\mu/m_t)$ [\,$\sim {\cal O}(\Gamma_{\!t}/m_t)$ in the (anti)top rest frame\,] and entails a symmetry of bHQET factorization with respect to changes of $v_{t,\bar t}^\mu$ and thus of $\vel$, which is one aspect of a more general class of symmetry transformations called reparametrization invariance~\cite{Luke:1992cs} that connects different orders in the bHQET $1/m_t$ expansion.

As a consequence, there is a power-suppressed freedom in the choice of the boost parameter $\vel$, which in the factorization theorem of~\eq{fact} appears in the
function $H^{(5,6)}_{\rm evol}$ and the momentum argument of the jet-function evolution factor $U_B^{(5)}$. The $m_t$ parameter appearing in $\vel$ is, however, not associated to any particular top mass renormalization scheme, but for the final evaluation it just has to be numerically chosen close to the invariant mass of the top quark in the resonance region. It is therefore not mandatory to re-expand the boost parameter $\vel$ when expressing the top quark mass in a short-distance renormalization scheme. The other consequence is, that variations $\vel\to\vel+\delta\vel$ with $\delta\vel\sim Q\Gamma_{\!t}/m_t^2$ in the peak region account for uncertainties due to the truncation of power corrections in the context of the leading power bHQET factorization. We conclude by noting that actually setting the parameter $m_t$ in $\vel$ equal to the pole mass artificially reintroduces the pole mass renormalon (albeit with formally power suppressed numerical effects) which cannot be cured by a re-expansion in terms of a short-distance mass, because there are no associated factorially divergent perturbative corrections in other parts of the factorization theorem. It is therefore not advisable to identify the parameter $m_t$ in $\vel$ as the pole mass.

The top mass parameter $m_t$ that appears in the argument of $H_m^{(6)}$ in \eq{Hevol} is identical to the mass used in the threshold decoupling relation of the strong coupling when transitioning between $6$ and $5$ flavors. This $m_t$ can be expressed in either the pole or $\MS$ scheme (with the same mass scheme employed in the $\alpha_s^{(6)}\!\! \leftrightarrow \! \alpha_s^{(5)}$ decoupling relation). Since $H_m$ only involves logarithms of the ratio $\mu_m/m_t$ (in addition to logs of $\vel$) it is also insensitive to the ${\cal O}(\Gamma_{\!t})$ fluctuations at the top threshold that are suppressed by ${\cal O}(\Gamma_{\!t}/m_t)$. The difference between the choice of pole or $\MS$ masses is accordingly found to be also numerically subleading.

\subsection{The Ultra-Collinear Sector}
\label{sec:ucollinear}
Upon integrating out the off-shell (small) components of the top quark field in bHQET, the leading-order dynamics of the remaining ultra-collinear fluctuations along the top and anti-top quark directions is captured by the 2-jettiness bHQET jet function, defined as
\begin{align}\label{eq:JBtau2}
& J_{B,\tau_2}^{(5)}\! (\shattau,\Gamma_{\!t},\delta m,\mu_B)
\\
& \qquad
=\!\!\int\! {\rm d} \hat s \, J_B^{(5)}\! (\hat s,\Gamma_{\!t},\delta m,\mu_B)\, J_B^{(5)} \!(\shattau-\hat s,\Gamma_{\!t},\delta m,\mu_B)
\,.\nn
\end{align}
Here ``ultra'' distinguishes the collinear modes in the peak region $\shattau \gsim 2 \Gamma_{\!t}$ [\,which are soft in the (anti)top quark rest frame\,] from higher virtuality collinear modes appearing in the tail region $\shattau\sim m_t$, where bHQET off-shellness power corrections become large and SCET provides the adequate description for collinear radiation. Here $J_B(\hat s_t, \delta m_t, \Gamma_{\!t}, \mu_B)$ is the familiar bHQET jet function that appears in di-hemisphere mass and jet mass distributions~\cite{Jain:2008gb,Fleming:2007xt,Hoang:2018zrp}. At leading order, the finite top width effects of the bHQET jet function $J_{B,\tau_2}^{(5)}\!(\shattau ,\Gamma_{\!t}, \delta m, \mu_B)$ can be expressed as a convolution of its stable-top version $J_{B,\tau_2}^{(5)}\!(\shattau,\Gamma_{\!t}=0 ,\delta m, \mu_B)$
with an inclusive Breit-Wigner function~\cite{Fleming:2007xt},
\begin{align}\label{eq:JBtau2stable}
J_{B,\tau_2}^{(5)}\!(\shattau ,\Gamma_{\!t} ,\delta m, \mu_B) &= \!\!\int \!\frac{{\rm d} \hat s^\prime}{\pi}
\frac{2\Gamma_{\!t}}{(2\Gamma_{\!t})^2 + (\shattau - \hat s^\prime)^2} \\
& \times J_{B,\tau_2}^{(5)}\!(\hat s^\prime ,\Gamma_{\!t}=0 , \delta m, \mu_B) \, , \nn
\end{align}
where the factor $2\Gamma_{\!t}$ arises from accounting for the widths from both top and antitop quarks.
This implies that the measurement on $t\tb$ final states is fully inclusive in the decay products as well as any radiation from them. The consistency of this inclusive treatment is ensured by considering the boosted limit $Q \gg m_t$ where the decay products from the top and anti-top quarks are collimated back-to-back along the direction of the thrust axis in distinct hemispheres. Both the boosted top decay products as well as gluon and light-quark radiation encoded in $J_B$ have typical angles $\sim 2m_t/Q$ relative to these axes.\footnote{Other global event shapes such as C-parameter are more sensitive to the kinematical distribution of the top decay products, such that the Breit-Wigner approximation is inaccurate even in the boosted limit~\cite{PreisserPhD}.} The variable $\hat s_{t}$ appearing in the (single) hemisphere jet functions $J_B^{(5)}$ in \eq{JBtau2} is equivalent to $\hat s_t = 2 v_t\cdot k$, where $k$ is the total residual momentum of the collimated system (after removing the contribution from the top quark mass). This captures the invariant mass of the decaying top quark system together with its ultra-collinear radiation, up to terms of ${\cal O}(\hat s_t^2/m_t)$~\cite{Fleming:2007qr}.
In the peak region, the ultracollinear fluctuations have virtuality of $\mathcal{O}(\hat s_{t,\bar t} \gsim \Gamma_{\!t})$ which leads to $\hat s_\tau =\hat s_t+\hat s_{\bar t}\gsim 2\Gamma_{\!t}$.
The natural choice for the renormalization scale of the 2-jettiness bHQET jet function is $\mu_B\sim \hat s_\tau \sim (Q^2\tau_2-2m_t^2)/m_t$.

For $\Gamma_{\!t}=0$ the 2-jettiness bHQET jet function $J_{B,\tau_2}^{(5)}$
has support only for non-negative $\hat s_\tau$ and equals $\delta(\hat s_\tau)/m_t$ at tree-level.
It is this threshold behavior which causes the strong top mass sensitivity of the $\tau_2$ distribution in the peak region. We emphasize, however, that while the partonic threshold is at $\hat s_\tau=0$, the observable peak position exhibited by the entire factorization theorem of \eq{fact} is determined coherently from the effects of the ultra-collinear and large-angle soft radiation together with the Breit-Wigner smearing. The well-known pole mass renormalon problem arises from higher order perturbative corrections in $J_{B,\tau_2}^{(5)}$ in the pole mass scheme, and is encoded in the size of the coefficients of plus distributions in $\hat s_\tau$. The pole mass renormalon can be remedied by using, instead of the pole mass $m_t^{\rm pole}$, a suitable short-distance mass scheme $m_t^{\rm sd}$ in the definition of $\hat s_\tau$ in Eq.~(\ref{eq:shatdef}). Because the bHQET jet functions are defined strictly at leading order in the $1/m_t$ expansion, switching the top mass renormalization scheme requires that one also accounts for the mass scheme correction
\begin{align}
\label{eq:generiglowscalemass}
\!\!\!\delta m(R) = m_t^{\rm pole}\! - m_t^{\rm sd}(R) \!=\! \sum_{i = 0} \biggl[\frac{\alpha_s^{(5)}(\mu_B)}{4\pi}\biggr]^i\delta m_i(R),
\end{align}
strictly to leading order in the $1/m_t$ expansion. This leads to the generic form
\begin{align}\label{eq:JbMom}
&J_{B,\tau_2}^{(5)}(\hat s_\tau, \Gamma_{\!t}=0,\delta m(R),\mu_B) \equiv \frac{1}{m_t^2}\sum_{i = 0} \biggl[\frac{\alpha_s^{(5)}(\mu_B)}{4\pi}\biggr]^i \nn \\
&\qquad \qquad\times\sum_{j = -1}^{2i+1}
\frac{B_{ij}}{\mu_B}{\cal L}_j\!\biggl(\frac{\hat s_\tau-4\delta m(R)}{\mu_B}\biggr) \, ,
\end{align}
where the $B_{ij}$ are constant coefficients.
This yields the 2-jettiness bHQET jet function for stable top quarks in an arbitrary (short-distance) mass scheme $m_t^{\rm sd}$, where it is {\it still strictly mandatory} to expand the dependence on $\delta m$ consistently in powers of the strong coupling $\alpha_s^{(5)}(\mu_B)$ such that the pole mass renormalon cancels order by order. Setting $\delta m=0$, one recovers the corresponding result in the pole mass scheme, which was calculated up to ${\cal O}(\alpha_s^2)$ in Refs.~\cite{Fleming:2007xt,Jain:2008gb}. Here, \mbox{${\cal L}_k(x) = \bigl[\Theta(x)\frac{\log^k(x)}{x}\bigr]_+$} is the standard plus function distribution with a vanishing integral over the range $x \in [0,1]$ for $k\geq 0$ and ${\cal L}_{-1}(x) = \delta(x)$.

From the expression in \eq{JbMom} we can also clearly see that the bHQET power counting requires that the perburbative series for the mass scheme correction $\delta m(R)$ obeys the scaling $\delta m\sim\hat s_\tau\ll m_t$. This shows from a power-counting point of view why low-scale short-distance masses have to be employed and the $\overline{\rm MS}$ mass, which has $\delta m\sim m_t$, is forbidden. Such low-scale short-distance masses always involve an infrared subtraction scale $R$, which is necessary to eliminate the
large ${\cal O}(\Lambda_{\rm QCD})$ renormalon corrections appearing in the pole mass scheme~\cite{Hoang:2008yj}.
To avoid upsetting the bHQET jet function power counting it is important that the series coefficients defining the low-scale short-distance mass in Eq.~(\ref{eq:generiglowscalemass}) have the property $\delta m_i(R)\propto R$, where the infrared scale $R$ is parametrically close to
the typical ultra-collinear scale, i.e.\ $R\sim\mu_B\sim\shattau$.
In our analysis we employ the MSR mass $m_t^{\rm MSR}(R)$~\cite{Hoang:2008yj,Hoang:2017suc,Hoang:2017btd,Mateu:2017hlz} that satisfies this requirement as explained in more detail in \sec{mass}.

We finally note that the overall factor $(1/m_t)^2$ appearing in the generic expression for the 2-jettiness bHQET jet function $J_{B,\tau_2}^{(5)}$ in \eq{JbMom} arises from \eq{JBtau2} since each $J_B$ has a $1/m_t$ factor~\cite{Fleming:2007xt}. The renomalization scheme dependence of this factor is power suppressed in the leading order bHQET expansion and therefore in principle not specified at the level of the factorization theorem of \eq{fact}.

\subsection{The Soft Sector and Nonperturbative Effects}
The thrust partonic soft function $\hat S_{\tau_2}^{(5)}$ accounts for the effects of large-angle soft radiation with respect to the thrust axis~\cite{Becher:2008cf,Hoang:2008fs,Chien:2010kc,Hornig:2011iu,Kelley:2011ng,Monni:2011gb}. In the context of inclusive observables, such as the global event shape \mbox{2-jettiness}, it accounts for the so-called ultrasoft modes, which have scale fluctuations of the order of $\mu_S \sim \shattau/\vel$, parametrically smaller than the typical scale $\mu_B$ for the ultra-collinear modes. Thus $\mu_S$ represents the smallest perturbative scale relevant for the $\tau_2$ distribution.

For global event shapes such as $\tau_2$ or jet-based observables without jet grooming, the leading nonperturbative effects from hadronization can be described via convolution with a shape function $F(k-2\Delta)$,
\begin{align}\label{eq:Sconvolution}
\!\!S_{\tau_2}(\ell,\mu_S) =\! \int_{0}^{\ell}\! \df k \, \hat S_{\tau_2}^{(5)}(\ell - k,\bar\delta=0, \mu_S) F(k - 2\Delta ),
\end{align}
where $F(k)$ has support for $k\geq 0$, peaks at \mbox{$k \sim \Lambda_{\rm QCD}$} \cite{Korchemsky:2000kp,Hoang:2007vb} and is by definition normalized to unity, $\int_0^\infty\df k F(k)=1$. Here, $\Delta$ is a model parameter which accounts for the minimum hadronic energy deposit in each hemisphere (hence the factor of 2), referred to as the `gap'. In the tail region where $\ell\sim\mu_S \gg \Lambda_{\rm QCD}$, one can expand for large $\ell$ and the most important nonperturbative effect is encoded in the first moment $\overline \Omega_1$ of the shape function,
\begin{align}\label{eq:O1bar}
S_{\tau_2}(\ell\gg\Lambda_{\rm QCD},\mu_S) = & \,\,\hat S_{\tau_2}^{(5)}(\ell,\bar\delta=0, \mu_S) \\ \nonumber
&
- 2\overline \Omega_1 \,\hat S_{\tau_2}^{(5)^\prime}(\ell,\bar\delta=0, \mu_S) +\ldots \,,\\
2\overline \Omega_1 \equiv &\,2\Delta + \!\int_0^\infty \!\df k \: k \, F(k) \, ,\nonumber
\end{align}
where
$\overline \Omega_1$ can also be expressed as a vacuum matrix element of soft Wilson lines~\cite{Lee:2006fn,Mateu:2012nk}, and the ellipses represent higher order terms in the expansion. The concrete form of the shape function we use for the generic numerical examination carried out in Sec.~\ref{sec:numerics} is provided in \app{model}.

As we move further into the peak region, $\mu_S$ decreases and eventually approaches the nonperturbative scale $\Lambda_{\rm QCD}$. Thus in the peak region $\mu_S \sim 1\,{\rm GeV} > \Lambda_{\rm QCD}$ and the effects of the shape function have to be accounted for exactly in terms of the convolution of \eq{Sconvolution}. Even though this in principle implies that an infinite amount of information could be required to fix the analytic form of the shape function, the normalization and the requirement that all moments of $F$ exist, together with the fact that factorization strictly demands convolution, allow us to reliably constrain the form of the shape function in terms of a few parameters by means of an expansion in optimally designed basis functions~\cite{Ligeti:2008ac}, see e.g.~\cite{Butenschoen:2016lpz} and \cite{Bernlochner:2020jlt}. In practice, determining the first moment of the shape function fixes the bulk of the information encoded in it for the whole $\tau_2$ spectrum.

The factorization into partonic soft and nonperturbative shape functions displayed in \eq{Sconvolution} depends on the regularization and renormalization schemes that are employed for the computatation of the partonic soft function. The argument `$\bar\delta=0$' shown in \eq{Sconvolution} stands for the standard $\overline{\rm MS}$ scheme. As was shown in Ref.~\cite{Hoang:2007vb}, these prescriptions entail that the partonic soft function has an ${\cal O}(\Lambda_{\rm QCD})$ renormalon which affects the partonic threshold at $\ell=0$ for the perturbative ultrasoft radiation~\cite{Gardi:2000yh,Hoang:2007vb} in a way very similar to how the ${\cal O}(\Lambda_{\rm QCD})$ pole mass renormalon affects the threshold $\hat s_\tau$ for the partonic ultra-collinear radiation. Both features are physically disentangled by the fact that the factorization theorem in \eq{fact} predicts that the ultrasoft effects on the hemisphere jet masses (and thus $\tau_2$) are enhanced by a factor $\vel$ compared to the ultra-collinear effects. This also entails that for a top quark mass determination, different c.m.\ energies $Q$ and simultaneous fits including parameters of the shape function must be considered to lift the degeneracy between $m_t$ and hadronization effects. This is in close analogy to the $\alpha_s$ determinations from $e^+e^-$ event-shape data carried out in~\Refs{Davison:2008vx,Abbate:2010xh,Hoang:2014wka}, where nonperturbative effects were not fixed from hadronization corrections in MC event generators, but from a simultanous fit using data obtained for different c.m.\ energies $Q$.

The difference between the ${\cal O}(\Lambda_{\rm QCD})$ renormalons affecting the partonic soft and bHQET jet functions is that the former cancels inside \eq{O1bar} with the nonperturbative matrix element $\overline \Omega_1$, while the latter is only an artifact of the pole mass and nonexistent when employing a short-distance mass scheme. One can remove the partonic soft function renormalon in an analagous way by re-expressing the first moment $\overline \Omega_1$ in a new scheme that includes a perturbative subtraction:
\begin{align}\label{eq:barDeltadef}
2\Omega_1(R_s) &\equiv \int_0^\infty \df k \: k \, F(k) + 2\overline \Delta(R_s)
,\\
\overline \Delta(R_s) &\equiv \Delta - \bar \delta(R_s) \, , \nn
\end{align}
where
\begin{align}\label{eq:delbarexpanded}
\bar \delta(R_s) =R_s \sum_{i = 0} \biggl[\frac{\alpha_s^{(5)}(\mu_S)}{4\pi}\biggr]^i\,\bar\delta_i\,,
\end{align}
is a perturbative series constructed such that it has exactly the same renormalon as $\overline \Omega_1$ and the $\overline{\rm MS}$ renormalized partonic soft function $\hat S_{\tau_2}^{(5)}(\ell,\bar\delta=0, \mu_S)$. \eq{barDeltadef} implies that
\begin{align}\label{eq:Omega1Rs}
\Omega_1(R_s) &=\overline \Omega_1 - \bar \delta(R_s) \\
&=\overline \Omega_1 - \Delta + \overline \Delta(R_s) \, . \nn
\end{align}
This entails the introduction of the scale $R_s \gtrsim \Lambda_{\rm QCD}$ (in analogy to the scale $R$ for the MSR mass dicussed below), which effectively represents an infrared cut for the partonic soft function which is then free from the ${\cal O}(\Lambda_{\rm QCD})$ renormalon ambiguity. This results in $\overline \Delta(R_s)$, and hence $\Omega_1(R_s)$, being renormalon free, and that instead of the $\overline{\rm MS}$ renormalized partonic soft function $\hat S_{\tau_2}^{(5)}(\ell ,\bar\delta=0, \mu_S)$, one employs the `gap subtracted' partonic soft function~\cite{Hoang:2007vb}:
\begin{align}\label{eq:shiftgap}
& \hat S_{\tau_2}(\ell, \bar \delta(R_s),\mu_S) \equiv \hat S_{\tau_2} (\ell - 2 \bar \delta(R_s), \mu_S) \\
& = \sum_{i = 0} \biggl[\frac{\alpha_s^{(5)}(\mu_S)}{4\pi}\biggr]^i\sum_{j = -1}^{2i+1}\frac{S_{ij}}{\mu_S}{\cal L}_j\!\biggl(\frac{\ell-2\bar \delta(R_s)}{\mu_S}\biggr) \, . \nn
\end{align}
In analogy to the partonic 2-jettiness bHQET jet function it is
{\it strictly mandatory} to expand the dependence on $\bar\delta$ consistently in powers of the strong coupling $\alpha_s^{(5)}(\mu_S)$ such that the soft function ${\cal O}(\Lambda_{\rm QCD}$) renormalon consistently cancels order by order. Furthermore,
to avoid upsetting the soft function power counting and to avoid the appearance of large logarithms in the subtraction it is mandatory that $R_s$ is parametrically close to
the typical soft scale, i.e.\ $R_s\sim\mu_S\sim\shattau/\vel$.
Setting $\bar\delta=0$ one recovers the $\overline{\rm MS}$ renormalized partonic soft function.
Hence, the hadron-level soft function becomes
\begin{align}\label{eq:gappedsoftfinal}
\!\!\!S_{\tau_2}(\ell,\mu) \!=\!\! \int_{0}^{\ell}\!\! \df k \, \hat S_{\tau_2}(\ell - k,\bar\delta(R_s), \mu) F(k - 2\overline \Delta(R_s) ).
\end{align}
We discuss the precise definition of the scheme that defines $\overline \Delta(R_s)$ in \sec{gap}.

At this point we note that there is a partial cancellation between the ${\cal O}(\Lambda_{\rm QCD})$ renormalons in the partonic soft and bHQET jet functions as the corresponding ambiguities and the associated diverging behavior of the perturbative series are equally severe but have an opposite sign~\cite{Hoang:2018zrp}. The extent of the cancellation is $Q$-dependent through the factor $\vel$ that enters the factorization convolution between the partonic soft and bHQET jet functions in \eq{fact}. As a consequence, the impact of the soft function renormalon in the 2-jettiness distribution increases with $Q$, while the impact of the jet function renormalon does not. This means that the overall effect of both ${\cal O}(\Lambda_{\rm QCD})$ renormalons may be hidden for a certain range of $Q$ values when the pole mass is used in the jet function and no gap subtraction is carried out for the partonic soft function. As we show in Sec.~\ref{sec:numerics} this indeed happens within the range of $Q$ values relevant top mass determinations. However, since simultaneous fits for different c.m.\ energies $Q$ are mandatory to independently determine the top mass and the parameters of the shape function without degeneracy, the impact of the individual ${\cal O}(\Lambda_{\rm QCD})$ renormalons cannot be avoided, so that the perturbative uncertainties in analyses accounting for both types of renormalon subtractions are systematically smaller than those when renormalon subtractions are not implemented.

\section{Top Mass Schemes}
\label{sec:mass}
In our analysis we employ two renormalization schemes for the top quark mass: the pole mass $m_t^{\rm pole}$ and the scale-dependent MSR mass $m_t^{{\rm MSR},(5)}(R)$~\cite{Hoang:2008yj,Hoang:2017suc,Hoang:2017btd}. The pole mass scheme has ---\,from a technical point of view\,--- the simplest implementation because we can set the residual mass term $\delta m$ appearing in \eqs{fact}{JbMom} to zero and all entries of $m_t$ discussed before to $m_t^{\rm pole}$. However, employing the pole mass scheme entails an ${\cal O}(\Lambda_{\rm QCD})$ renormalon ambiguity, which visibly destabilizes the order-by-order behavior of the cross section in the peak region as we show explicitly in Sec.~\ref{sec:numerics}. This can be systematically avoided by using a suitable short-distance mass scheme. The MSR mass $m_t^{{\rm MSR},(5)}(R)$ is a short-distance scheme that is derived from the $\overline{\rm MS}$ mass and represents an extension of the $\overline{\rm MS}$ mass concept: While the $\overline{\rm MS}$ mass $\overline m_t^{(6)}(\mu)$ is suitable for scales $\mu\ge m_t$, the MSR mass is appropriate for scales $R\le m_t$.

For the $\overline{\rm MS}$ mass we will use the notation $\overline m_t^{(6)}\equiv\overline m_t^{(6)}(\mu=\overline m_t^{(6)})$ below.
In the approximation that all quark flavors lighter than the top are massless (which we adopt in our analysis), the defining series for
$m_t^{\rm pole} - m_t^{{\rm MSR},(5)}(R)$ is obtained\footnote{Here we employ the scheme that was called `natural MSR' mass in Ref.~\cite{Hoang:2017suc}.} from the corresponding series for $m_t^{\rm pole} - \overline m_t^{(6)}$ by removing all corrections arising from the self-energy diagrams with top quark loops and by setting $\overline m_t^{(6)}\to R$ as well as $\alpha_s^{(6)}\to\alpha_s^{(5)}$:
\begin{align}\label{eq:poleMSRdef}
\!\!\!\!\delta m(R) = m_t^{\rm pole} \!- m_t^{{\rm MSR},(5)}\!(R) \!=\! R\sum_{i=1} \biggl[\frac{\alpha_s^{(5)}(R)}{4\pi}\biggr]^i a_i .
\end{align}
We give the details of the MSR scheme definition and the numerical values of the coefficients in \app{msr}.
This means that the MSR mass is a scheme derived from the $\overline{\rm MS}$ mass, but where the virtual off-shell fluctuations in the on-shell self-energy from scales beyond $R$ (which includes virtual top quark effects) are integrated out. The MSR mass is therefore designed for top mass dependent observables sensitive to soft QCD dynamics.

The scale $R$ can be interpreted as the resolution scale below which virtual self-energy and real ultra-collinear radiation is treated as unresolved, so that only self-energy contributions above $R$ are absorbed into the mass.
This means that $m_t^{{\rm MSR},(5)}(R)$ for \mbox{$R<m_t$} and $\mbar_t^{(6)}(R)$ for \mbox{$R>m_t$} contain self-energy contributions coming only from scales above $R$. This interpretation entails that in the limit $R\to 0$, where all virtual self-energy and real ultra-collinear radiation is treated as resolved and all virtual self-energy contributions are absorbed in the mass, we approach the pole mass, which is precisely expressed in \eq{poleMSRdef}. The renormalon ambiguity of the pole mass can thus be seen to be associated with the problem that the limit $R\to 0$ involves crossing the Landau pole of the strong coupling which a priori cannot be carried out in an unambiguous way.

Here we use the interpretation in~\cite{Hoang:2017suc}, where the top MSR mass $m_t^{{\rm MSR},(5)}(R)$ is regarded as the 5-flavor extension of the \mbox{6-flavor} $\overline{\rm MS}$ mass $\overline m_t^{(6)}(R)$ for scales $R$ below the top quark mass and where both mass schemes are matched at the scale \mbox{$R=\overline m_t^{(6)}$}. The matching relation is given in \eq{MSRnMSbarmatch}. For scales $R< \mbar_t$ the MSR mass evolves with the \mbox{$R$-evolution} equation
\begin{align}\label{eq:RRGE}
\!\!\frac{\df}{\df\log (R)}m_t^{{\rm MSR},(5)}(R) =
- R\!\sum_{ n = 0} \gamma^R_n \, \biggl[\frac{\alpha_s^{(5)}(R)}{4\pi}\biggr]^{n+1}\!,
\end{align}
where $\gamma^R_n$'s are obtained from the coefficients $a_i$'s in \eq{poleMSRdef} using the procedure outlined in \app{Revolution}.

As explained in \sec{ucollinear}, the consistent use of a short-distance mass in \eq{fact} entails that the MSR mass scale $R$ satisfies the parametric relation $R\sim\mu_B\sim\hat s_\tau$, which means that $R$ depends on $\tau_2$ and $m_t^{{\rm MSR},(5)}$ adopts the status of a dynamical scale-dependent `mass coupling' in complete analogy to the well-known concept of the scale- and flavor-number-dependent strong coupling $\alpha_s$. This dynamical treatment of the top quark MSR scheme resums important large logarithms via the \mbox{$R$-evolution} equation~\eq{RRGE}. The reader should note that the RHS of the \mbox{$R$-evolution} equation is linear in $R$, which differs from the common logarithmic renormalization group equations. This linear evolution is an essential aspect of properly treating the physical mass effects that govern the resonance/close-to-mass-shell dynamics of heavy colored particles.

In our numerical analysis we use $\overline m_t^{(6)}=\overline m_t^{(6)}(\overline m_t^{(6)})$ as the standard reference mass value, which we quote as our main input and from which we then calculate the MSR or $\overline{\rm MS}$ masses at the respective scales needed within the factorization formula in \eq{fact}.
For the flavor-number-dependent-strong coupling $\alpha_s^{(5,6)}(\mu)$ we always use matching at \mbox{4-loops} \cite{Chetyrkin:1997un,Chetyrkin:2005ia,Schroder:2005hy} and running at 4-loops~\cite{Larin:1993tp,Czakon:2004bu}, where the flavor matching is carried out at $\mu=\overline m_t^{(6)}$.

We finally note that the analytic properties of the pole mass ${\cal O}(\Lambda_{\rm QCD})$ renormalon in terms of knowledge on the large-order behavior of the perturbation series obtained in the pole mass scheme are by now very well understood, see e.g.\ Refs.~\cite{Beneke:2016cbu,Hoang:2017suc} for the case where all quarks except for the top are assumed massless and Ref.~\cite{Hoang:2017btd} where finite bottom and charm masses are included systematically.\footnote{This knowledge implies a more precise understanding of the size of the renormalon ambiguity, but not that the ambiguity itself is eliminated.} One of the most interesting observations in this context is that the large-order asymptotic behavior (for some unknown reason) universally sets in at ${\cal O}(\alpha_s)$ and is already well saturated at ${\cal O}(\alpha_s^2)$, which is one of the reasons why the pole mass renormalon has received significant attention in the literature. Another useful (but also confusing) consequence of this fact is that is it very easy to devise different types of low-scale short-distance masses from either physical~\cite{Czarnecki:1997sz,Beneke:1998rk,Hoang:1998ng,Hoang:1998hm,Hoang:1999ye,Fleming:2007qr,Jain:2008gb} or conceptual considerations~\cite{Pineda:2001zq,Hoang:2008yj,Hoang:2017suc,Hoang:2017btd}. All are ---\,as long as the resolution scale $R$ is assigned appropriate values\,--- similarly effective in minimizing mass-related QCD corrections and stabilizing the perturbation series already at ${\cal O}(\alpha_s^{1,2})$. The MSR mass $m_t^{{\rm MSR},(5)}(R)$ provides a unifying concept to connect all low-scale short-distance masses with the $\overline{\rm MS}$ mass via its renormalization group equation given in~\eq{RRGE}.

\section{Soft Gap Subtraction}
\label{sec:gap}
The ${\cal O}(\Lambda_{\rm QCD})$ renormalon of the soft function $\hat S_{\tau_2}^{(5)}$ has large-order properties very similar to the ${\cal O}(\Lambda_{\rm QCD})$ pole mass renormalon in the bHQET jet function. It is known to differ from the latter just due to a different normalization~\cite{Hoang:2007vb} and the sign difference already mentioned at the end of Sec.~\ref{sec:fact}. However, compared to the pole mass renormalon, the soft function ${\cal O}(\Lambda_{\rm QCD})$ renormalon is harder to pinpoint quantitatively at low orders due to the soft function anomalous dimension and hadron mass effects~\cite{Salam:2001bd,Mateu:2012nk}. Therefore, the soft function renormalon does not seem to exhibit the same universality as the pole mass renormalon (even though it is phenomenologically equally relevant). Apart from $e^+e^-$ event shapes~\cite{Lee:2006nr}, it is still largely unknown whether or in which way the ${\cal O}(\Lambda_{\rm QCD})$ renormalon in $\hat S_{\tau_2}^{(5)}$ appears in a universal manner in soft functions relevant for collider observables which are linearly sensitive to large-angle soft radiation.

It has so far been the practice to define the renormalon subtraction series $\bar{\delta}(R_s)$ in \eq{delbarexpanded} directly from the soft function, see Refs.~\cite{Hoang:2007vb,Hoang:2008fs} for such approaches. In our analysis we employ a modified version of the prescription suggested in Ref.~\cite{Hoang:2008fs}. It is based on the Fourier transform of the renormalon-subtracted partonic soft function
\begin{align}
\label{eq:Spositionspace}
\tilde S_{\tau_2}^{(5)}(y,\bar\delta(R_s),\mu_S) = & \int\! \df\ell \: e^{-i y \ell } \hat S_{\tau_2}^{(5)}(\ell, \bar \delta(R_s), \mu_S)\\
=&\,\tilde S_{\tau_2}^{(5)}(y,\bar\delta=0,\mu_S) \,e^{-2i\bar{\delta}(R_s) y}\,,\nn
\end{align}
where the gap subtraction series $\bar\delta(R_s)$ is factored into the exponential factor shown in the second line. It is therefore possible to define an expression for $\bar\delta(R_s)$ which cancels the ${\cal O}(\Lambda_{\rm QCD})$ renormalon of the soft function by the condition
\begin{align}
\label{eq:gapdef}
\bar{\delta}( R_s ) &\equiv
\frac{R_s}{2}
\log \biggl[\tilde S_{\tau_2}^{(5)}\biggl(\frac{1 }{i R_s }, \bar\delta=0, R_s\biggr)\biggr]
\\
&=
\frac{R_s }{2}
\sum_{i = 1} \biggl[\frac{\alpha^{(5)}_s(R_s)}{4\pi}\biggr]^i\sum_{j = 0}^{i+1} s_{ij}\, \gamma_E^j
\, ,
\nn
\end{align}
where $s_{ij}$ are coefficients of the fixed-order series expansion of $\log[\tilde S_{\tau_2}^{(5)}(y, \mu)]$ shown below explicitly in \eq{lnSBcoeffs}. When the gap subtraction series $\bar\delta(R_s)$ is used in the factorization theorem, it is crucial that the renormalization scale of the strong coupling $\alpha_s(R_s)$ is re-expressed in terms of $\alpha_s(\mu_S)$, the coupling used in the series for the soft function, to ensure a systematic order-by-order cancellation of the renormalon. This is detailed in \app{gap}.

This definition of the gap subtraction can be contrasted with the one used in \Ref{Hoang:2008fs} where the subtraction series was instead related to a derivative of the soft function logarithm:
\begin{equation}\label{eq:gapdef1}
\!\!\bar \delta^{\tiny \text{\,Ref.\cite{Hoang:2008fs}}} (\mu_S, R_s) \!
\equiv\!
\frac{R_s e^{\gamma_E}}{2}
\frac{\df\!\log \tilde S_{\tau_2}^{(5)}(y,\mu_S)}{\df \!\log (iy )}
\bigg|_{iy e^{\gamma_E}=\frac{1}{R_s}}\!
.
\end{equation}
In this definition, the scale of the strong coupling is $\mu_S$ by construction, and the gap subtraction inherits a non-trivial anomalous dimension in $\mu_S$ from the soft function. In \app{gap} we describe a set of generic gap subtraction schemes that include \eqs{gapdef}{gapdef1} as special cases.

While both definitions in \eqs{gapdef}{gapdef1} are perfectly viable subtraction schemes (i.e.\ equally effective at asymptotic large orders), the series $\bar \delta^{\,\tiny \text{Ref.\cite{Hoang:2008fs}}}$ in \eq{gapdef1} is numerically zero at $\mathcal{O}(\alpha_s)$ for $\mu_S = R_s$ because the one-loop non-cusp anomalous dimension vanishes, $\gamma^{S_{\tau}}_0 = 0$. This necessitates choosing $R_s$ strictly below $\mu_S$ in the peak region to reduce the size of the ${\cal O}(\alpha_s)$ correction. This can, however, be problematic when considering ${\cal O}(\alpha_s^2)$ corrections because in the peak region the soft scale \mbox{$\mu_S\sim \hat s_\tau/\vel$} is already parametrically smaller than the top quark width such that setting $R_s>\mu_S$ can lead to instabilities. On the other hand, for the gap subtraction definition in \eq{gapdef}, we have $s_{10} = 15.053$ at one loop which, in addition to being non-zero, is also numerically sizable allowing for the implementation of an effective gap subtraction for the more natural setting $R_s = \mu_S$. We will therefore adopt the gap subtraction scheme defined in Eq.~(\ref{eq:gapdef}) in our analysis.

The $R$-evolution of the gap parameter corresponding to the scheme defined by $\bar \delta$ in \eq{gapdef} is given by
\begin{align}\label{eq:RsRGE}
&\overline \Delta (R_1) - \overline \Delta (R_0)
=
\\
&\qquad -\sum_{n = 0}^\infty
\bar \gamma_n^R \! \int_{R_0}^{R_1}\! \df R
\biggl[\frac{\alpha_s^{(5)}(R)}{4\pi}\biggr]^{n+1}
\, ,
\nn
\end{align}
where the anomalous dimension coefficients $\bar \gamma_n^R$ are derived from the fixed-order coefficients $s_{ij}$ in \eq{gapdef} following the steps laid out in \app{Revolution}.

In our numerical analysis we
take the first moment value for $\Omega_1^{\tiny \text{\,Ref.\cite{Hoang:2008fs}}}(2\, {\rm GeV})$ obtained in \Ref{Abbate:2010xh} as the input and determine from it $\Omega_1(2\, {\rm GeV})$ in the gap scheme defined in Eq.~(\ref{eq:gapdef}) as well as $\overline \Omega_1$ for the case of no gap subtraction. For the no-gap case we use \mbox{$\Delta= 0.1$} GeV as the input value of the gap parameter in \eq{O1bar} which fixes the analytic form of the shape function $F(k)$. From this we calculate $\overline \Delta (R_s=2\, {\rm GeV})$ in the scheme of Eq.~(\ref{eq:gapdef}) using \eq{barDeltadef}.
This input then unambiguously fixes the value of $\overline\Delta(R_s)$ at any scale $R_s$ using the \mbox{$R$-evolution} equation \eq{RsRGE}, thus determining the form of the gap-subtracted shape function \mbox{$F(k-2\overline\Delta(R_s))$} in \eq{gappedsoftfinal}. For the $R$-evolution in \eq{RsRGE} we employ 2-loop precision~\cite{Hoang:2008fs,Monni:2011gb,Kelley:2011ng}. For $F(k)$ we adopt the parametrization given in Ref.~\cite{Ligeti:2008ac}, see App.~\ref{app:model} for explicit expressions. We note that this approach implies that the form of the shape function $F(k-2\overline\Delta(R_s))$ entering the factorization theorem~\eq{fact} depends dynamically on the value of the physical 2-jettiness variable $\tau_2$. In the peak region this dependence is, however, quite weak because the scale $\mu_S$ saturates, see Sec.~\ref{sec:profile}.

\section{N$\mathbf{^3}$LL resummed cross section}
\label{sec:N3LLresult}
For the numerical evaluation of the factorization formula in \eq{fact},
we find it convenient to work in the position (Fourier) space where the convolutions involving the bHQET jet and and soft functions and their respective renormalization evolution factors become simple products.
(The results in this section could equally well be expressed using the Laplace transform.)

\subsection{Stable-top cross section without renormalon subtractions}

We first discuss the stable top quark cross section for the case of having neither soft function gap nor mass subtractions.
The Fourier transforms of the stable top \mbox{2-jettiness} bHQET jet and soft functions are defined via
\begin{align}
\label{eq:JbHQETpos}
J_{B,\tau_2}^{(5)}(\hat s_\tau,\mu) &=\! \int\! \frac{\df x}{2\pi} \: e^{i x \hat s_\tau } \tilde J^{(5)}_{B,\tau_2} (x, \mu) \, , \\
\hat S_{\tau_2}^{(5)}(\ell^+, \mu) &= \!\int\! \frac{\df y}{2\pi} \: e^{i y \ell^+ } \tilde S_{\tau_2}^{(5)}(y,\mu)\nn \,,
\end{align}
where for brevity we have dropped the zero arguments $\delta m=0$, $\bar\delta=0$ and $\Gamma_{\!t}=0$.
Thus the position space jet and soft functions have the form:
\begin{align}\label{eq:lnSBcoeffs}
&m_t^2\tilde J^{(5)}_{B,\tau_2} (x, \mu_B) = \\
&\qquad \exp \Biggl\{ \sum_{i = 1} \biggl[\frac{\alpha^{(5)}_s(\mu_{B})}{4\pi}\biggr]^i \
\sum_{j = 0}^{i+1} b_{ij} \log^j\bigl(ie^{\gamma_E} x \mu_B \bigr) \Biggr\}, \nn\\
& \tilde S^{(5)}_{\tau_2}(y, \mu_S) = \nn \\
&\qquad \exp \Biggl\{\sum_{i = 1} \biggl[\frac{\alpha^{(5)}_s(\mu_S)}{4\pi}\biggr]^i\,\sum_{j = 0}^{i+1} s_{ij} \log^j\bigl(ie^{\gamma_E} y \mu_S\bigr)\Biggr\}
\, , \nn
\end{align}
such that \eq{fact} can be written as
\begin{align}
\label{eq:factposUnSub}
&\frac{{\rm d}\hat \sigma^{\Gamma_{\!t} \rightarrow 0} }{{\rm d}\tau_2}(\hat s_\tau)=\sigma_0 m_t\, Q^2 \!\!\int \!\frac{\df x}{2\pi} \: e^{i x \shattau}\\
&\qquad\times H^{(5,6)}_{\rm evol}\! (Q, m_t, \vel, \mu ; \mu_H, \mu_m) e^{K_B^{ (5)}(\mu, \mu_B) + K_S^{ (5)}(\mu, \mu_B)} \nn\\
&\qquad\times
\bigl(i e^{\gamma_E} x \mu_B\bigr)^{\omega^{(5)}_B(\mu,\mu_B)}
\bigl(i e^{\gamma_E} \vel x \mu_S\bigr)^{\omega^{(5)}_S(\mu,\mu_S)}
\nn\\
& \qquad \times \tilde J_{B,\tau_2}^{(5)}(x, \mu_B)\, \tilde S^{(5)}_{\tau_2}(\vel x,\mu_S) \,,\nn
\end{align}
where due the convolution in \eq{fact} the soft function in position space is evaluated at $y = \vel x$ and the evolution kernels $K_i^{(n_f)}$ and $\omega_i^{(n_f)}$'s are defined in App.~\ref{app:formulae}. The RG-evolved hard factor in \eq{factposUnSub} is given by
\begin{align}\label{eq:HevolRG}
&H^{(5,6)}_{\rm evol} \!(Q,m_t, \vel, \mu; \mu_H, \mu_m) \equiv \\
& \quad H_Q^{(6)}\!(Q,\mu_H)\, H_m^{(6)}(m_t, \vel, \mu_m)\, e^{K_{H_Q}^{(6)}(\mu_m, \mu_H) + K_{v}^{\gamma(5)}(\mu_m, \mu)} \nn \\
& \quad \times
\Bigl(\frac{\mu_H}{Q}\Bigr)^{\!\omega^{(6)}_{H_Q}\!(\mu_m, \mu_H)} \! \vel^{-\omega^{(5)}_{v}\!(\mu_m, \mu)}\,.\nn
\end{align}
Here $\mu$ is the common final renormalization scale of all the RG evolution factors. Taking the inverse Fourier transform back to distribution space we find
\begin{align}\label{eq:factpos0}
&\frac{{\rm d}\hat \sigma^{\Gamma_{\!t} = 0} }{{\rm d}\tau_2}(\hat s_\tau)
=
\frac{{\rm d}\hat \sigma^{(0)} (\hat s_\tau, \partial_\Omega) }{{\rm d}\tau_2}
\frac{e^{\gamma_E \Omega}}{\Gamma(-\Omega)}
\bigg|_{\Omega = \tilde \omega^{(5)} (\mu_S, \mu_B)}
\, ,
\end{align}
where for later convenience we have defined the following function of the derivative operator $\partial_\Omega$:
\begin{align}
\label{eq:factpos1}
&\frac{{\rm d}\hat \sigma^{(0)} }{{\rm d}\tau_2}(\hat s_\tau, \partial_\Omega)
\equiv \sigma_0\frac{m_tQ^2}{\shattau}
H^{(5,6)}_{\rm evol}\! (Q, m_t, \vel, \mu ; \mu_H, \mu_m)\nn \\
&\times e^{K_B^{ (5)}(\mu, \mu_B) + K_S^{ (5)}(\mu, \mu_B)}
\Bigl(\frac{\mu_B}{\shattau}\Bigr)^{\!\omega^{(5)}_B(\mu,\mu_B)} \!
\Bigl(\frac{\vel \mu_S}{\shattau}\Bigr)^{\!\omega^{(5)}_S(\mu,\mu_S)}
\nn\\
& \times \tilde J_{B,\tau_2}^{(5)}\biggl[ \partial_\Omega + \log\Bigl(\frac{\mu_B}{\shattau}\Bigr)\biggr]
\tilde S^{(5)}_{\tau_2}\bigg[ \partial_\Omega + \log\Bigl(\frac{\vel \mu_S}{\shattau}\Bigr)\biggr]
\,.
\end{align}
It acts on the function of $\Omega$ shown in \eq{factpos0}, and the outcome is evaluated at the following $\mu$-independent evolution kernel between the bHQET jet and soft scales:
\begin{align}\label{eq:tildeOmega}
\tilde \omega^{(5)} (\mu_S, \mu_B) \equiv \omega_S^{(5)} (\mu, \mu_S) + \omega_B^{(5)}(\mu, \mu_B) \, .
\end{align}

We note that since the jet scale is always above the soft scale, one has $\tilde \omega^{(5)} < 0$. In Eq.~(\ref{eq:factpos1}) the arguments of the position-space jet and soft functions are understood to replace the corresponding logarithms shown in Eqs.~(\ref{eq:lnSBcoeffs}). For sake of brevity we have suppressed the arguments $\mu_B$ and $\mu_S$ that appear in the running coupling, as shown in \eq{lnSBcoeffs}. The meaning of the superscript `$(0)$' on $\hat \sigma$ will be clarified below in Eq.~(\ref{eq:sigmaDelOmega}). We also note that the dependence of the result in \eq{factpos1} on $\hat s_\tau$ is defined in terms of rational power plus-distributions which have support for $\hat s_\tau\ge 0$ in the case of stable top quarks~\cite{Fleming:2007xt}. To account for the fixed-order corrections contained in the product of the functions $H_Q^{(6)}$, $H_m^{(6)}$, $J_{B,\tau_2}^{(5)}$ and $\tilde S^{(5)}_{\tau_2}$ at N$^k$LO we expand their product strictly to ${\cal O}(\alpha_s^{k})$. The relevant formulae for the evolution kernels and their Fourier transforms are presented in \app{formulae}. We collect the numerical results for the anomalous dimensions in \app{anomDim} and the fixed-order expressions for all the factorization functions $H_Q^{(6)}$, $H_m^{(6)}$, $J_{B,\tau_2}^{(5)}$ and $\tilde S^{(5)}_{\tau_2}$ up to NNLO in \app{fixedOrder}.

Finally,
since we carry out resummation at the level of the differential cross section, when implementing the cross section at N$^k$LL+$\mathcal{O}(\alpha_s^{k-1})$ accuracy for $k\ge 1$ (referred to as `unprimed' orders), we explicitly incorporate the $\mathcal{O}(\alpha_s^{k})$ plus-function boundary condition \cite{Almeida:2014uva} in order to correctly sum up logarithms that are counted as N$^k$LL in the exponent of the cumulative distribution. This amounts to including the {\it single logarithmic} terms appearing at $\mathcal{O}(\alpha_s^{k})$ in the jet and soft functions. For the `primed' orders N$^k$LL$'$, or equivalently, N$^k$LL\,+\,$\mathcal{O}(\alpha_s^k)$ accuracy, this is not necessary as the ${\cal O}(\alpha_s^k)$ fixed-order matching already includes this single logarithmic term. The loop-order of the theoretical ingredients for the primed and unprimed orders are summarized in Tab.~\ref{tab:ordercounting}. We refer to Ref.~\cite{Abbate:2010xh} for further details on primed and unprimed orders.
\begin{table}[t!]
\begin{tabular}{c|ccccccc}
& ~cusp~ & ~non-cusp~ & ~matching~ & ~$\beta[\alpha_s]$~ & ~~$\gamma_R$~~ & ~~$\delta$~~ \\
\hline
LL & 1 & - & tree & 1 & - & - \\
NLL & 2 & 1 & tree & 2 & 1 & - \\
N$^2$LL & 3 & 2 & 1 & 3 & 2 & 1 \\
N$^3$LL & 4 & 3 & 2 & 4 & 3 & 2 \\
\hline
NLL$^\prime$ & 2 & 1 & 1 & 2 & 1 & 1 \\
N$^2$LL$^\prime$ & 3 & 2 & 2 & 3 & 2 & 2 \\
\end{tabular}
\caption{Loop corrections required for specified orders. In the last two columns $\gamma_R$ and $\delta$ refer to either soft or MSR-mass subtractions.}
\label{tab:ordercounting}
\end{table}

\subsection{Renormalon subtractions}
\label{sec:MSRrenormalon}

We now describe how the renormalon subtractions with respect to the top quark mass and soft function gap are to be included starting from the unsubtracted stable top cross section in \eq{factpos0}. First we recall that the $\delta m$ dependence in the bHQET jet function $J_{B,\tau_2}^{(5)}$ results from re-expressing the pole mass $m_t^{\rm pole}$ contained in $\shattau$ in terms of the MSR mass. From \eq{shatdef} we have
\begin{align}\label{eq:shatexpanded}
\shattau&= \frac{Q^2 \tau_2 - 2 \bigl[m_t^{\rm MSR}(R) + \delta m(R)\bigr]^2}{m_t^{\rm MSR}(R) + \delta m(R)} \\
&= \shattau^{\rm MSR} (R) - 4 \delta m(R) + {\cal O} \Bigl(\frac{\alpha_s \Gamma_{\!t}}{m_t}\Bigr)
\,, \nn
\end{align}
where
\begin{align}
\shattau^{\rm MSR} (R) \equiv \frac{Q^2 \tau_2 - 2 \bigl[m_t^{\rm MSR} (R)\bigr]^2}{m_t^{\rm MSR}(R)} \, .
\end{align}
The two terms shown in the second line of \eq{shatexpanded} represent those to be accounted for in \eq{fact} since power-suppressed contributions in the peak region must be systematically dropped for consistency. For the term $\delta m(R)$, which contains the pole mass renormalon ambiguity, this is particularly important to achieve the order-by-order cancellation of the pole mass renormalon.
Thus, the stable-top bHQET jet function using \eq{JbMom} in position space can be expressed up to NNLO as
\begin{align}
\label{eq:JbHQETposSub}
&\tilde J_{B,\tau_2}^{(5)}(x,\delta m,\mu_B) = \int \df \hat s^\prime \: e^{-i x \hat s'} J^{(5)}_{B,\tau_2} (\hat s'- 4\delta m, \mu_B) \nn \\
&\quad= \biggl[1
-(ix) 4\delta m
+ (ix)^2\frac{( 4\delta m)^2}{2!}
\biggr]J^{(5)}_{B,\tau_2} (x, \mu_B) \,,
\end{align}
where we have dropped the zero argument $\Gamma_{\!t}=0$ in the jet function and the argument $R$ in the mass subtraction $\delta m(R)$ for simplicity. We have kept only terms at most quadratic in $\delta m(R)$ so that the pole mass renormalon can be consistently canceled to ${\cal O}(\alpha_s^2)$. Likewise, the soft gap subtraction can be incorporated using \eq{Spositionspace} such that
\begin{align}\label{eq:SposSub}
&\tilde S_{\tau_2}^{(5)}(\vel x, \bar \delta, \mu_S) = e^{-2i\vel \bar \delta x} \tilde S_{\tau_2}^{(5)}(\vel x, \mu_S)
\\
&\qquad = \biggl[1
-(ix) 2\vel \bar \delta
+ (ix)^2\frac{( 2\vel \bar \delta )^2}{2!}
\biggr]\tilde S_{\tau_2}^{(5)}(\vel x, \mu_S) \,,\nn
\end{align}
where we dropped the argument $R_s$ in the gap subtraction $\bar\delta(R_s)$.
Including the subtraction terms in \eqs{JbHQETposSub}{SposSub} and {\it strictly} expanding to ${\cal O}(\alpha_s^2)$ we arrive at the following expression for the renormalon-subtracted cross section for stable top quarks:
\begin{align}\label{eq:factposSub}
&\frac{{\rm d}\hat \sigma^{\Gamma_{\!t} = 0} }{{\rm d}\tau_2}\bigl(\hat s_\tau, \delta m(R), \bar \delta(R_s )\bigr)
\\
&\ =
\sum_{n= 0}^{2} \frac{{\rm d}\hat \sigma^{(n)} }{{\rm d}\tau_2}(\hat s^{\rm MSR}_\tau(R), R_s ,\partial_\Omega)
\frac{e^{\gamma_E \Omega}}{\Gamma(-\Omega)}
\biggr|_{\Omega = \tilde \omega^{(5)} (\mu_S, \mu_B) + n}
\,.\nn
\end{align}
Here we have
\begin{align}
\label{eq:sigmaDelOmega}
&\frac{{\rm d}\hat \sigma^{(n)} }{{\rm d}\tau_2} (\hat s^{\rm MSR}_\tau(R),R_s, \partial_\Omega)
\\
&\qquad =\frac{(-1)^n}{n!} \biggl[\frac{\delta_{\rm tot}(R, R_s)}{\shattau^{\rm MSR}(R)}\biggr]^n \frac{{\rm d}\hat \sigma^{(0)} ( \hat s^{\rm MSR}_\tau(R), \partial_\Omega) }{{\rm d}\tau_2} \, ,
\nn
\end{align}
where the total subtraction series $\delta_{\rm tot}$ has the form
\begin{align}
\label{eq:totalsub}
\delta_{\rm tot}(R, R_s) \equiv 4 \delta_m(R ) +2 \vel \,\bar \delta(R_s) \,,
\end{align}
and ${\rm d}\hat \sigma^{(0)}/{\rm d}\tau_2$ is given in Eq.~(\ref{eq:factpos1}).
It is essential to consistently drop terms of ${\cal O}(\alpha_s^3)$ and higher (concerning the fixed-order corrections in the hard, jet and soft functions as well as the renormalon subtractions) in the product in \eq{sigmaDelOmega}. Note that the derivatives in the sum over $n$ in \eq{factposSub} are evaluated at $\Omega = \tilde \omega + n$. For simplicity we will display the functions $\hat s_\tau^{\rm MSR}, \delta_m, \bar \delta$ and $\delta_{\rm tot}$ without their arguments $R$ and $R_s$ below.

\subsection{Including the top width}
From \eq{JBtau2stable} we see that the cross section for unstable top quarks involves an additional Breit-Wigner convolution such that\footnote{The analytical results shown in this section were first derived in Ref.~\cite{Butenschoen:2016lpz}.}
\begin{align}\label{eq:factunstable0}
\frac{{\rm d}\hat \sigma}{{\rm d}\tau_2}(\hat s^{\rm MSR}_\tau, \delta m, \bar \delta )
&=
\int_0^\infty \!\frac{{\rm d} \hat s^\prime}{\pi}
\frac{2\Gamma_{\!t}}{(2\Gamma_{\!t})^2 + (\shattau^{\rm MSR} - \hat s^\prime)^2} \\
&\qquad \times
\frac{{\rm d}\hat \sigma^{\Gamma_{\!t} = 0} }{{\rm d}\tau_2}(\hat s^\prime, \delta m, \bar \delta ) \, ,\nn
\end{align}
where as a result of the stable top cross section, the integration is bounded below by $\hat s' = 0$. However, $\shattau^{\rm MSR}$ (or $\shattau$ in the pole mass scheme) for the unstable top 2-jettiness cross section can be negative as well.
To incorporate this convolution in \eq{factposSub} we first note that the cross section is proportional to $1/(\hat s')^{\tilde \omega^{(5)} + 1}$, as can be seen from \eq{factpos1} using \eq{tildeOmega}.
This factor can be brought to the right of the $\partial_\Omega$ derivatives using the identity
\begin{align}
&\frac{1}{(\hat s')^{\Omega + 1}} \biggl[ \partial_\Omega + \log \Bigl(\frac{\mu_B}{\hat s'}\Bigr)\biggr]^n
=\frac{1}{\bigl(\sqrt{\shattau^2 + (2\Gamma_{\!t})^2}\,\bigr)^{\Omega + 1}}
\\
&
\times \biggl[ \partial_\Omega + \log \biggl(\frac{\mu_B}{\sqrt{\shattau^2 + (2\Gamma_{\!t})^2}}\biggr)\biggr]^n\!
\biggl(\frac{\sqrt{\shattau^2 + (2\Gamma_{\!t})^2}}{\hat s'}\,\biggr)^{\! \Omega + 1} \,,
\nn
\end{align}
where we remind the reader that $\hat s'$ is the integration variable in \eq{factunstable0} whereas $\shattau$ (or $\shattau^{\rm MSR}$ in the MSR mass scheme) is related to the $\tau_2$ measurement as defined in \eq{shatdef}.
The analogous relation also holds for the $\partial_\Omega$ derivatives associated to the soft function. Hence, we need to evaluate the following convolution:
\begin{align}\label{eq:calIdef}
{\cal I}\bigl(\Omega, \shattau, 2\Gamma_{\!t}\bigr) &\equiv
\frac{1}{\Gamma(- \Omega)}\!
\int_0^\infty \!
\frac{\df \hat s^\prime}{\pi}\frac{2\Gamma_{\!t}}{(2\Gamma_{\!t})^2 + (\shattau - \hat s^\prime)^2}
\\
&\qquad \times
\biggl(\frac{\sqrt{\shattau^2 + (2\Gamma_{\!t})^2}}{\hat s'}\,\biggr)^{\! \Omega + 1}
\nn \\
&=
\frac{\phi\bigl(\frac{\shattau}{2\Gamma_{\!t}}\bigr) \,\Gamma(2 + \Omega )}{\Gamma\bigl(1 + (1+\Omega)\phi\bigr)\Gamma\bigl(1 - (1+\Omega)\phi\bigr)}
\, ,\nn
\end{align}
which has a smooth $\Omega\to0$ limit, and where we have defined
\begin{align}
\phi(x) \equiv \frac{1}{2} + \frac{1}{\pi} \!\arctan(x)
\, .
\end{align}
In the limit $\Gamma_{\!t}\to0$ one smoothly recovers the stable-top results. For $\Gamma_{\!t} = 0$ we find that ${\cal I} = 0$ when $\shattau < 0$.
Using these expressions in \eqs{factposSub}{factunstable0} we now arrive at the final expression for the unstable-top cross section with renormalon subtractions:
\begin{align}\label{eq:factunstable1}
&\frac{{\rm d}\hat \sigma }{{\rm d}\tau_2}(\hat s^{\rm MSR} _\tau, \delta m, \bar \delta\, )
=
\sum_{n= 0}^{2} \frac{{\rm d}\hat \sigma^{(n)} }{{\rm d}\tau_2}\!\Bigl(\!\sqrt{(\shattau^{\rm MSR})^2 + (2\Gamma_{\!t})^2}, \partial_\Omega\!\Bigr)\nn
\\
&\qquad \qquad\quad
\times e^{\gamma_E \Omega}{\cal I}\bigl(\Omega, \shattau, 2\Gamma_{\!t}\bigr)
\Bigr|_{\Omega = \tilde \omega^{(5)} (\mu_S, \mu_B) + n}
\, .
\end{align}
Finally, the hadron-level cross section is obtained from the partonic cross section via convolution with the nonperturbative shape function:
\begin{equation}
\frac{{\rm d} \sigma}{{\rm d}\tau_2}(\tau_2) = \!\int\! \df k\: \frac{{\rm d}\hat \sigma}{{\rm d}\tau_2} (\shattau^{\rm MSR}(R) - \vel k) \, F(k - 2\overline\Delta(R_s) )\,.
\end{equation}

\section{Profile Functions}
\label{sec:profile}

To properly sum large logarithms we use \mbox{$\tau_2$-dependent} renormalization scales $\mu_i(\tau_2)$, $R(\tau_2)$ and $R_s(\tau_2)$, called profile functions~\cite{Ligeti:2008ac,Abbate:2010xh}. They have canonical scaling in the resummation regions, and freeze at a perturbative scale in the resonance region to avoid the breakdown of perturbation theory for anomalous dimensions. In the far tail region, they become equal to the hard scale to reproduce the fixed-order perturbative expansion with a common scale $\mu$. They are expressed in terms of $7$ parameters which can be varied to estimate perturbative uncertainties. Following Refs.~\cite{Butenschoen:2016lpz,Dehnadi:2016snl} we employ a natural generalization of the profile functions devised for massless event shapes in~\cite{Hoang:2014wka}, to which they reduce in the massless limit.

\begin{figure}[t!]
\includegraphics[width = 0.44\textwidth]{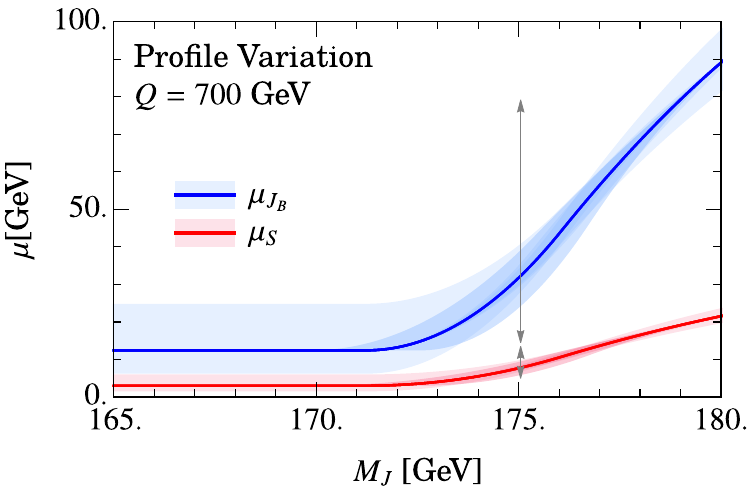}
\caption{\label{fig:prof} Examples of variation of the jet and soft scales to estimate perturbative uncertainty. The $e_{J_0}$ parameter is related to a `trumpet' variation in the jet scale that turns off in the fixed-order region. For the soft scale we vary $\mu_0$, the perturbative value that the soft scale freezes to in the nonperturbative region.}
\end{figure}

The strategy to estimate perturbative uncertainties involves varying all the profile functions up and down by at most a factor of $2$ and $1/2$, respectively, including a shape dependent variation in the jet scale, as well as varying the value at which the soft scale freezes in the nonperturbative region. We show bands for the latter two variations in \fig{prof}, and indicate the factor of $2$ variations by arrows. We scan over these profile variations by generating a sample of 500 profiles were all their parameters are varied simultaneously with random choices within their allowed ranges (see Ref.~\cite{Abbate:2010xh} for details on this general approach).
The concrete form of the profile functions and how their parameters are varied are given below.
The total uncertainty is determined by the envelope of the resulting cross sections.

For the hard and mass matching scales we use $\tau_2$-independent functions $\mu_H= e_H Q$ and $\mu_m= \sqrt{e_H}\, m_t$, respectively, which depend on the center-of-mass energy and top quark mass, as well as on a free parameter $e_H$, which has the default value $1$ and in our random scan is varied in the interval $[0.5,2]$. The variation of the two matching scales are correlated to keep the correct hierarchies. For the top quark mass parameter $m_t$, which appears in the matching scale $\mu_m$ as well as in other variables entering the profile functions discussed below, we adopt the numerical value $m_t=m^{\rm MSR}(R=5\,{\rm GeV})$.

Our $\tau_2$-dependent profile functions are implemented through the soft scale $\mu_S(\tau_2)$, which in the peak region is parametrized with the following piecewise function
\begin{align}\label{eq:profile-soft}
\!\!\mu_{S}\!
=\!\! \left\{\!\!\begin{array}{l r}
e_{\rm np}\,\mu_0\,, & \!\!\!\! \!\!\!\!\! \! \tau_2^{\rm min} \!\le \! \tau_2 \! \le\! t_0 \\[0.15cm]
\zeta[e_{\rm np}\,\mu_0, \mu_S(\tau_2>t_1),t_0, t_1,\tau_2]\,,&\!\!\!\!\!\! t_0 \!< \tau_2\! <\! t_1 \\[0.15cm]
\! \!\Bigl( 1 \!+\! \frac{n_s e_{\rm slope}}{n_s + \hat m_t-\tau_2^{\rm min}}\Bigr) r_{\!s}\mu_H(\tau_2-\tau_2^{\rm min}), & \!\!\! t_1 \!\le \!\tau_2 \!<\! t_2
\end{array}
\right.\!\!\!,\!\!\! \nn\\
\end{align}
where $\tau_2^{\rm min}=2\hat{m}_t^2$ with $\hat m_t\equiv m_t/Q$.
We refer to the three corresponding intervals as the ``non-perturbative", ``bHQET'', and \mbox{``SCET-resummation''} regions. The
function $\zeta[f_a(\tau_2), f_b(\tau_2), t_a, t_b,\tau_2]$ smoothly connects any two linear functions $f_{a,b}(\tau_2)$ that end/begin at the points $t_{a,b}$. This is achieved by means of two quadratic polynomials of $\tau_2$ smoothly joined at $\tau_2=(t_a+t_b)/2$, where the explicit formula can be found in Ref.~\cite{Hoang:2014wka}. The parameter $n_s$ has the default value $0.5$ and is varied in the range $|n_s-0.5|\leq 0.025$. Its effect in the peak region is relatively mild. The default slope in the SCET-resummation region is set by $r_s = 2$ and using the default value $0$ for the variable $e_{\rm slope}$. Slope variations are implemented by varying $e_{\rm slope}$ in the interval $[1/1.13-1,1.13-1]$. Note that the rescaling factor to the left of $r_s$ approaches $1$ in the massless limit. In a similar way, the parameter affecting the flat non-perturbative region is $e_{\rm np}$. Its default value is $1$ and it is varied in the interval $[1/\sqrt{2},\sqrt{2}]$ with $\mu_0=3$\,GeV. The values of the transition points $t_{0,1}(m_t, Q)$ depend on the mass and center-of-mass energy and have the form
\begin{align}\label{eq2:transition}
t_0=\,& \frac{2}{(Q/1 {\rm GeV})} + \frac{d_0}{(Q/1 {\rm GeV})^{0.5}} + \tau_2^{\rm min} \,, \\
t_1=\,& \frac{2.25}{(Q/1 {\rm GeV})^{0.75}} + \frac{d_1}{(Q/1 {\rm GeV})^{0.5}} + \tau_2^{\rm min} \,,\nn\\
t_2 =\,& n_2 + \hat{m}_t\,,\nn
\end{align}
with $|d_{0,1}| \leq 0.05$ and \mbox{$|n_2-0.25| \leq 0.025$}. Their default values are $d_{0,1}=0.05$ and $n_2=0.25$.
For the energies and masses considered in this article one has $t_0<t_1<t_2$. For the jet scale profile function $\mu_J(\tau_2)$ we first define $\tilde\mu_J(\tau_2)=\sqrt{e_H}\,\mu_S(\tau_2)/\hat m_t$ and $t_s=n_s+\hat m_t$, and then use the piecewise function
\begin{equation}\label{eq:profile-B}
\!\!\mu_{J}\!
=\! \left\{\!\!\begin{array}{l r}
[1 + \tilde e_J (t_0-t_s)^2]\,\tilde\mu_J(t_0)\,, & \! \! \!\!\! \tau_2^{\rm min} \!\le \tau_2 \!\le \! t_0 \\[0.15cm]
\zeta[\mu_J(t_0),\mu_J(\tau_2 > t_1),t_0, t_1,\tau_2]\,,&\! \!\!\!\!t_0\! < \tau_2\! <\! t_1 \\[0.15cm]
[1 + \tilde e_J (\tau_2-t_s)^2]\,\tilde\mu_J(\tau_2)\,, & \!\!\! \!\! t_1\! \le \!\tau_2\! < t_2
\end{array}
\right.\!\!\!.\!\!
\end{equation}
Here the jet-function parameter $\tilde e_J$ is defined with a rescaling factor
\begin{equation}\label{eq:rescalingf}
\tilde e_J = e_J \biggl[\frac{n_s-(t_0 - \tau_2^{\rm min}) }{t_s - t_0}\biggr]^{2}\,,
\end{equation}
with variations $|e_J|\leq 1.5$ used for assessing uncertainties and the default value $e_J=0$. For $m_t=0$ we recover $e_J= \tilde e_J$. We set the mass and soft-function renormalon subtraction scales to their respective canonical values: $R_s(\tau_2)=\mu_S(\tau_2)$ and $R(\tau_2)=\mu_J(\tau_2)$.

\section{Numerical Analysis}
\label{sec:numerics}

In this section we present a numerical analysis of the bHQET N$^3$LL\,+\,$\mathcal{O}(\alpha_s^2)$ 2-jettiness peak region cross section based on the factorization formula of \eq{fact}. We remind the reader that this factorization theorem is based on the bHQET limit and does not account for subleading terms related to higher powers of $\hat s_\tau$ (kinematic power corrections) and $m_t/Q$ (mass power corrections). As was shown in Refs.~\cite{Butenschoen:2016lpz,Dehnadi:2016snl,PreisserPhD}, these corrections are formally power-suppressed in the peak region, but still important numerically for a realistic phenomenological analysis concerning the top mass dependence of the \mbox{2-jettiness} peak-region line shape.
In the following we therefore carry out a generic numerical analysis pointing out important features of the bHQET 2-jettiness peak region cross section at this order related to the convergence of the perturbative series, as well as the impact of the jet and soft function renormalons and the improvement related to their subtractions. A phenomenological analysis aiming for a systematic study of other sources of theoretical uncertainties is postponed to future work.

For our analysis we have two independent codes to implement all cross sections, one in \texttt{fortran}~\cite{gfortran} and one in \texttt{C++} using SCETlib~\cite{SCETlib}. Numerical integrations are carried out using either \texttt{quadpak}~\cite{quadpack} or the \texttt{gsl} library~\cite{contributors-gsl-gnu-2010}.

Before getting to results some comments on our parametric input are in order.
For the subsequent discussion we use $\alpha_s^{(5)}(m_Z) = 0.118$ with $m_Z = 91.1876\,\mbox{GeV}$ as input for the strong coupling. As input for the top quark mass we take the standard $\overline{\rm MS}$ mass $\overline m_t^{(6)}\equiv\overline m_t^{(6)}(\overline m_t^{(6)}) =160$\,GeV. The conversion (matching) to the MSR top mass at the scale $R=\overline m_t^{(6)}$ is based on the formulae derived in Ref.~\cite{Hoang:2017suc} employed at ${\cal O}(\alpha_s^2)$ and given in \eqs{DeltaAn}{anAll}. The evolution of the MSR top mass to the renormalon subtraction profile scale $R$ is obtained via \eq{RRGE} using the $\mathcal{O}(\alpha_s^3)$ R-anomalous dimensions, such that
\begin{align}
\label{eq:MSR2GeV}
m_t^{\rm MSR}(R=2\,\mbox{GeV}) = 169.537\,{\rm GeV}\,,
\end{align}
for the MSR top mass at $2$\,GeV, which one can interpret as the (renormalon-free) kinematic mass that governs the top-mass dependence of the peak position. The same top mass value is used in the boost parameter $\vel$ defined in \eq{velDef} when the MSR mass scheme is used. For comparison we will also discuss results for the cross section in the pole mass scheme and without gap subtraction.
At this point one has to recall that the pole mass is, due to its renormalon ambiguity, an order-dependent concept, where the size of the fixed-order corrections in its relation to a short-distance mass at a given order depends on the renormalization scale of the short-distance mass. Thus to achieve comparable theoretical predictions employing the MSR and the pole mass schemes (i.e.\ with peak positions that are compatible), it is essential to apply fixed-order conversion from the MSR to the pole mass at the renormalization scale of the MSR mass that is employed in the peak region of the distribution (which is the part of the distribution that carries the highest top quark mass sensitivity). Furthermore the order of conversion has to match the fixed-order input used for the theoretical calculation~\cite{Hoang:2020iah,Hoang:2017btd}. In the peak region of the distribution the appropriate renormalization scale is just the bHQET jet function scale $\mu_J$, around $10$\,GeV, see Sec.~\ref{sec:profile}.
Thus, the proper way to determine the pole mass is to use the fixed-order conversion from $m_t^{\rm MSR}(10\,\mbox{GeV})$ as was pointed out in \Ref{Butenschoen:2016lpz}.
Therefore, to determine the pole mass for the following analysis we first determine $m_t^{\rm MSR}(10\,\mbox{GeV})$ using Eq.~(\ref{eq:MSR2GeV}) and R-evolution and then apply fixed-order conversion to the pole scheme at the scale $R=10$\,GeV. As we are employing ${\cal O}(\alpha_s^2)$ fixed-order corrections to the bHQET jet function at the highest N$^3$LL order of our analysis, this fixed order conversion must be carried out at 2-loops. Using this procedure, we find
\begin{align}\label{eq:mtPole}
m_t^{\rm pole}= 169.718\,\mbox{GeV} \, .
\end{align}
This input value for the top quark pole mass ensures that at N$^3$LL the peak position
in the pole mass scheme is compatible with that obtained in the MSR mass scheme.

Lastly, in order to fix the form of the nonperturbative model function $F(k)$, see Eq.~(\ref{eq:Sconvolution}) and \app{model}, we have to specify values for the first moment $\Omega_1$. In analogy to the top quark mass there are also different schemes for $\Omega_1$ related to the definition of the gap subtraction parameter. Thus the corresponding values for $\Omega_1$ also have to be fixed with some care. Here we aim to adopt values for $\Omega_1$ consistent with the $e^+e^-$ thrust analysis of Ref.~\cite{Abbate:2010xh}, using the fact that the same universal soft function that is given in Eq.~(\ref{eq:Sconvolution}) also appears in the thrust factorization theorem for massless quark production. The value of $\Omega_1$ determined in \Ref{Abbate:2010xh}, $\Omega_1^{\tiny \text{\,Ref.\cite{Hoang:2008fs}}}(R_s=2\, {\rm GeV})= 0.323 \pm 0.045\,{\rm GeV}$, was based on the gap subtraction scheme suggested in \Ref{Hoang:2008fs} [explained in detail in and below Eq.~(\ref{eq:gapdef1})]. For our analysis it needs to be converted to the gap scheme of Eq.~(\ref{eq:gapdef}) adopted in this work, referred to as $\Omega_1(R_s)$, as well as to the unsubtracted $\MS$ gap scheme $\overline \Omega_1$ (which still contains the soft function renormalon). For these conversions we must use the 2-loop fixed-order formulas in \eqs{Omega1OldNewDiff}{Omega1Rs2GeV}, respectively, since at the highest N$^3$LL order of our analysis we employ the soft function at ${\cal O}(\alpha_s^2)$. This gives
\begin{align}\label{eq:Omega1s}
\Omega_1(R_s = 2\,\mbox{\rm GeV}) &= 0.739\, {\rm GeV}
\,,\\
\overline \Omega_1 &= 0.276 \,{\rm GeV}
\nn \,,
\end{align}
where we convert at the scale $R_s=2$\,GeV, the typical value of the soft function profile in the peak region.
Fixing $\Delta = 0.1\,{\rm GeV}$ for the unsubtracted gap [\,see Eq.~\eqref{eq:O1bar}\,] and using \eq{barDeltadef} we obtain \mbox{$\overline \Delta (R_s = 2\, {\rm GeV}) = 0.563$\,GeV} for the gap term entering Eq.~\eqref{eq:gappedsoftfinal}. With this choice of parameters we can use the same shape function parameters as employed in \Ref{Abbate:2010xh}, which corresponds to taking $\lambda= 0.349\,{\rm GeV}$ and \mbox{$c_2 = 0.05$} in \eq{basis1}. We then
use \eq{RsRGE} with ${\cal O}(\alpha_s^2)$ running to determine $\overline \Delta (R_s)$ at the $\tau_2$-dependent $R_s$ values required by the gap subtraction profile function.

\begin{figure}[t]
\includegraphics[width = 0.48\textwidth]{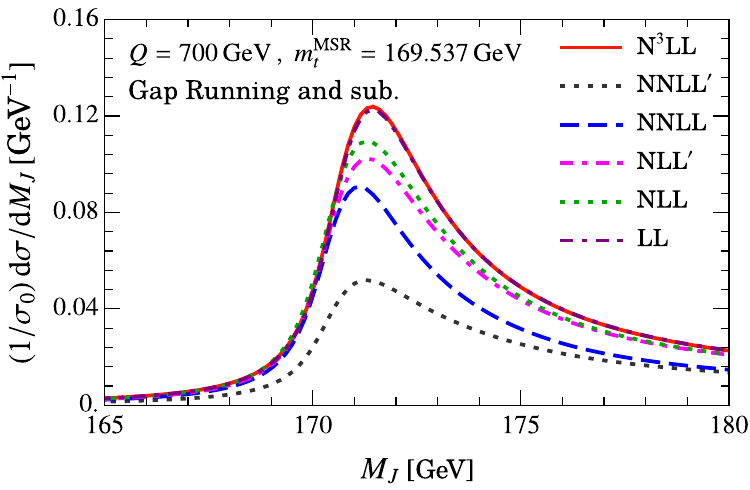}
\caption{\label{fig:renorm-corr} Perturbative convergence of the 2-jettiness cross section, with all curves normalized with the Born cross section $\sigma_0$. The peak location shows excellent convergence, while the normalization corrections are significant and exhibit slower convergence.}
\end{figure}

In Fig.~\ref{fig:renorm-corr} the 2-jettiness differential cross section in the MSR mass scheme with gap-subtraction
is shown for $Q=700$\,GeV as a function of the inclusive jet mass variable $M_J$ [\,see \eq{MJdef}\,] at all primed and unprimed orders up to N$^3$LL for the default set of profile functions (see Sec.~\ref{sec:profile}). Here the primed orders include all contributions of the unprimed with the addition of the fixed-order matrix elements at one higher order in $\alpha_s$. Our results at NNLL$^\prime$ and N$^3$LL are new and have not been analyzed before in the literature.

In Fig.~\ref{fig:renorm-corr} all curves are normalized to the Born-level massless total cross section $\sigma_0$. The behavior of the curves at the different orders therefore reflects the effects of the perturbative corrections to the shape as well as to the normalization of the cross section. We see that, apart from sizable normalization corrections which happen to be positive for all subsequent orders, the convergence with respect to the peak location and the shape is excellent. From the difference in the curves obtained from primed and unprimed orders we can also see that the effects on the normalization from higher order corrections in the renormalization group equations (and thus the resummation of large logarithmic terms) are smaller than those in the fixed-order matrix elements.

\begin{figure}[t!]
\begin{center}
\hspace{-0.15cm}\includegraphics[width = 0.465\textwidth]{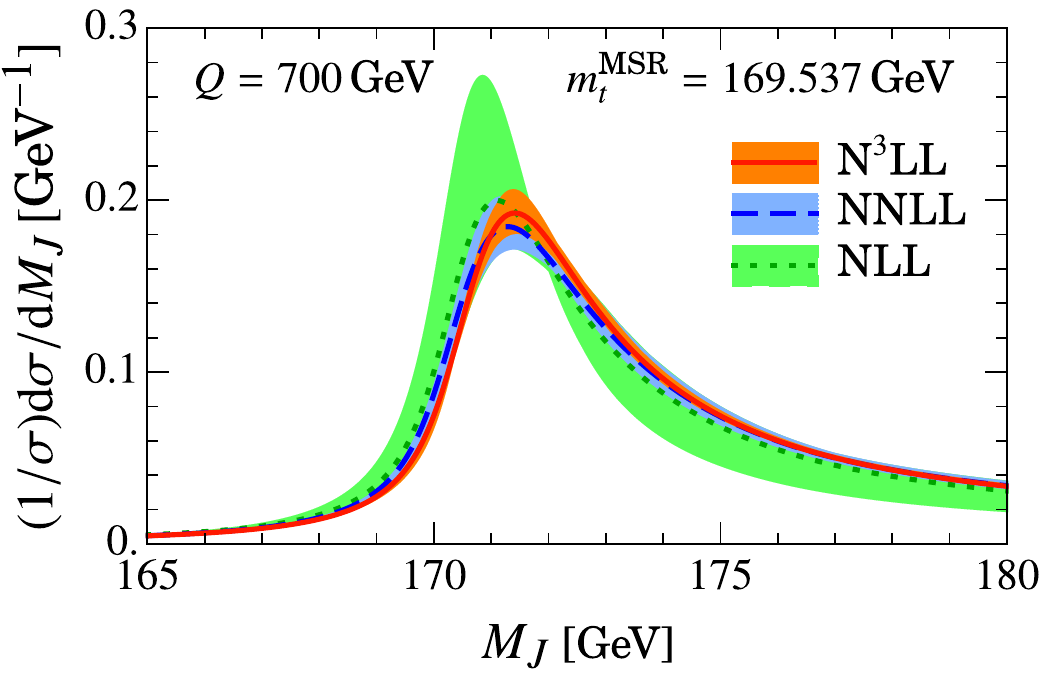}
\hspace{0.15cm}\includegraphics[width = 0.45\textwidth]{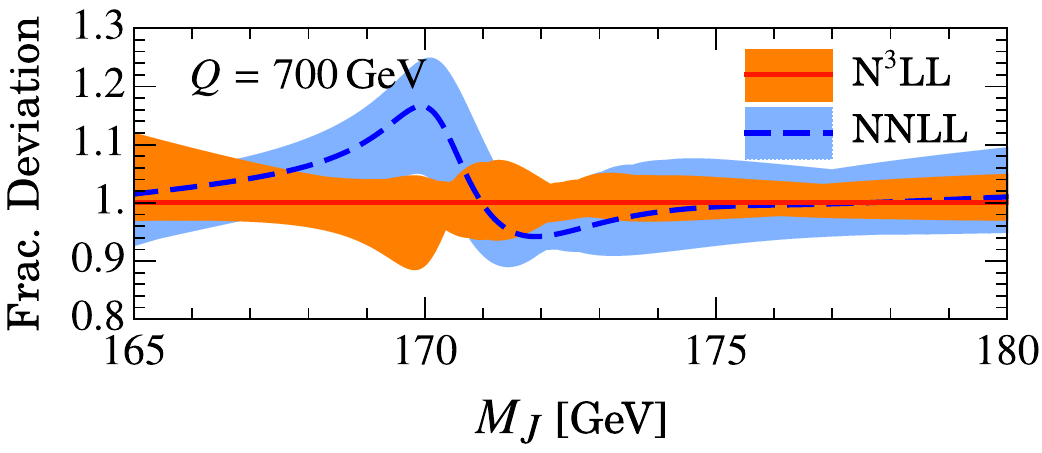}
\vspace{-0.2cm}
\hspace{-0.25cm}\includegraphics[width = 0.465\textwidth]{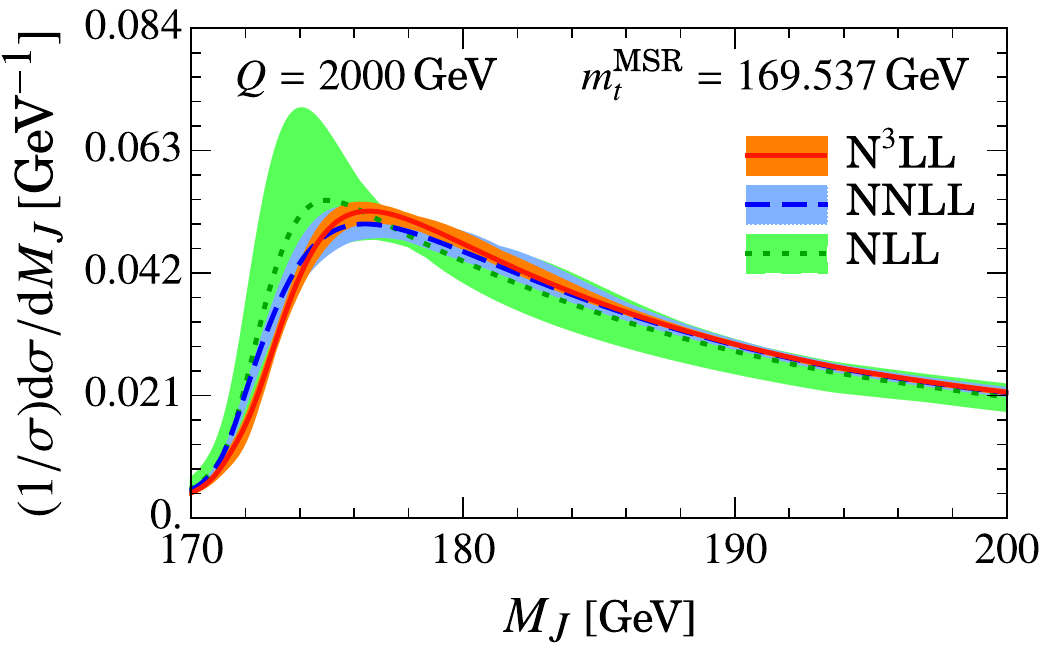}
\vspace{-0.3cm}
\hspace{0.1cm}\includegraphics[width = 0.45\textwidth]{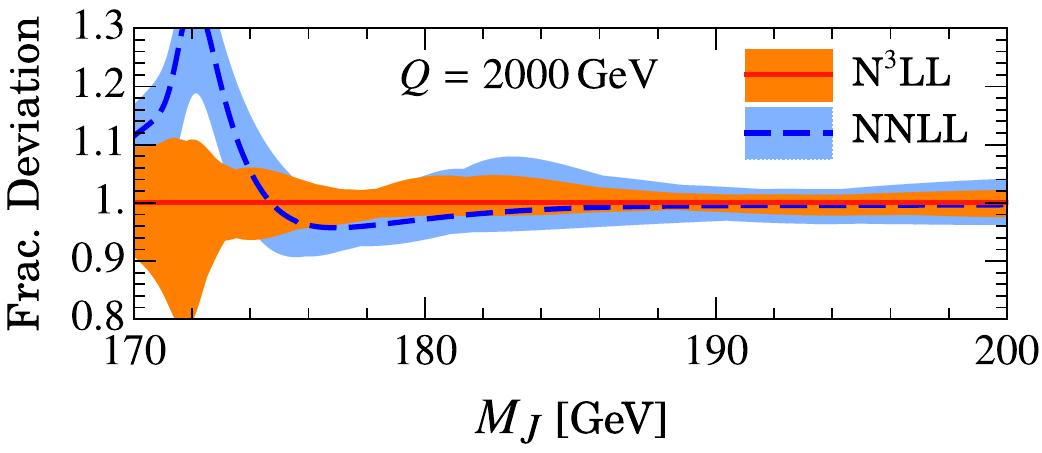}
\vspace{-0.3cm}
\end{center}
\caption{\label{fig:profile-MSR-mu0sqrt2} Perturbative convergence and uncertainty bands for self-normalized cross sections at $Q=700\,{\rm GeV}$ (upper panels) and $Q=2000\,{\rm GeV}$ (lower panels) in the MSR mass scheme and with gap subtractions. All curves are normalized over the displayed ranges. The two smaller panels show the same results at the two highest orders, but as a fractional deviation from the central N$^3$LL result.}
\end{figure}

In order to analyse the effects of the higher-order corrections to the distribution shape and its order-dependent perturbative uncertainty, it is useful to normalize the curves from the different orders to a common $M_J$ interval. In Fig.~\ref{fig:profile-MSR-mu0sqrt2} the 2-jettiness differential cross sections at $Q=700$\,GeV (upper panels) and \mbox{$Q=2000$\,GeV} (lower panels) are shown in the MSR mass scheme with gap subtractions using default profile functions. The results are normalized to the $M_J$ interval displayed in the respective panels at NLL (green dotted line), NNLL (blue dash-dotted line) and N$^3$LL (red solid line). Primed orders are not displayed to avoid cluttering. We also display uncertainty bands with the corresponding colors at each of these three orders. These bands are derived by determining the upper and lower value of the distributions (for each $M_J$ value) obtained by considering $500$ profile functions generated randomly within the profile function parameter ranges given in Sec.~\ref{sec:profile}. To generate the bands, each cross section from a given profile is normalized to the displayed $M_J$ range. The central curves exhibit excellent perturbative convergence for the shape. The width of each band illustrates the size of the perturbative uncertainty, which nicely decreases with increasing order. For better visibility the error bands and lines are displayed once more in the lower parts of each plot showing the fractional deviation from the central N$^3$LL curve.
At $Q=700$\,GeV for $M_J\ge 171\,{\rm GeV}$ the relative uncertainty in the peak region is $\pm(4$--$10)$\% at NNLL and \mbox{$\pm (3$--$7)$\%} at N$^3$LL.
In contrast, at $Q=2000$\,GeV\, for $M_J\ge 175\,{\rm GeV}$ the relative uncertainty in the peak region is $\pm(3$--$8)\%$ at NNLL and $\pm(1$--$5)\%$ at N$^3$LL.

\begin{figure}[t!]
\begin{center}
\hspace{-0.15cm}\includegraphics[width = 0.465\textwidth]{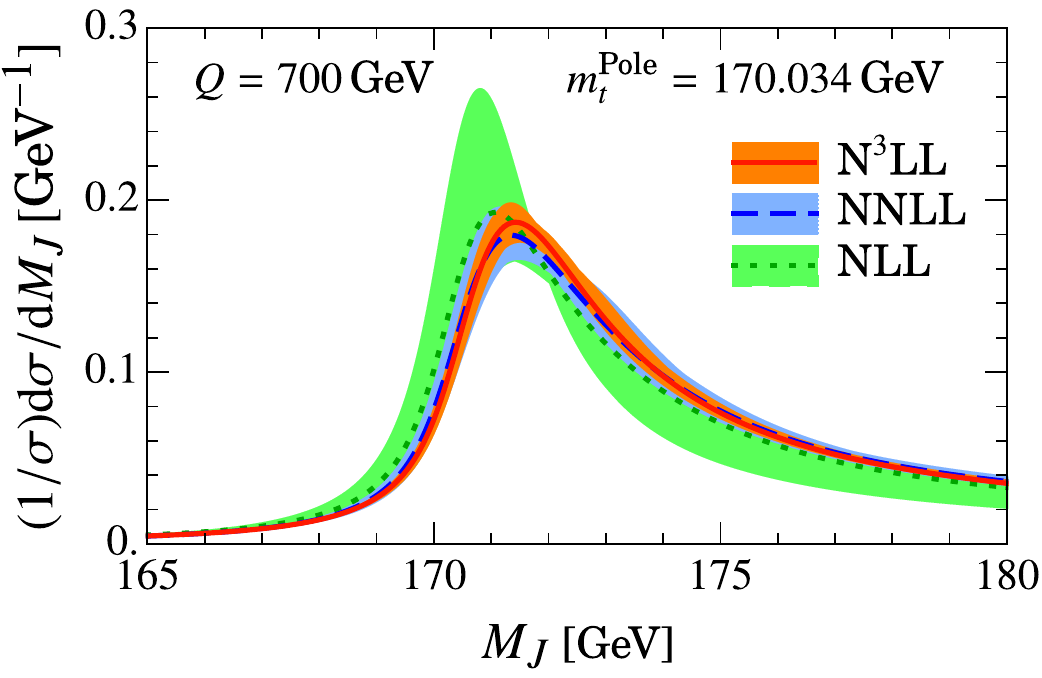}
\hspace{0.15cm}\includegraphics[width = 0.45\textwidth]{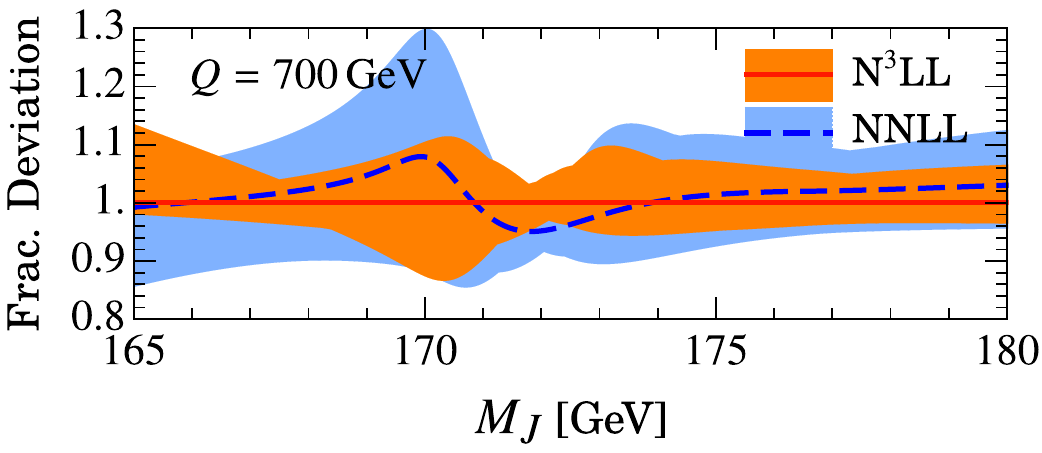}
\vspace{-0.2cm}
\hspace{-0.25cm}\includegraphics[width = 0.465\textwidth]{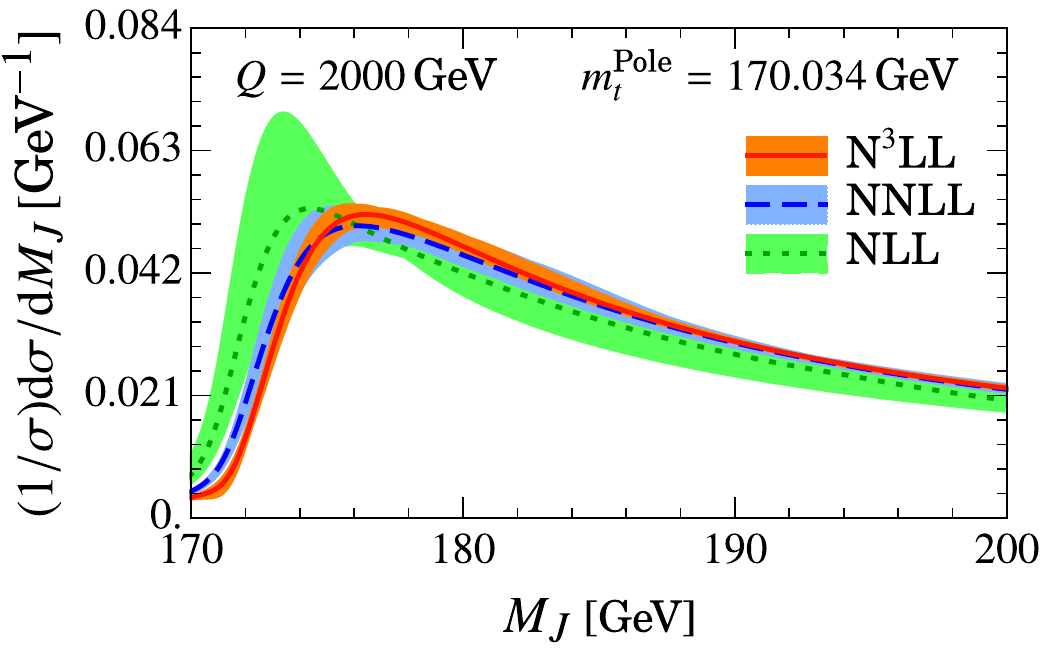}
\vspace{-0.3cm}
\hspace{0.1cm}\includegraphics[width = 0.45\textwidth]{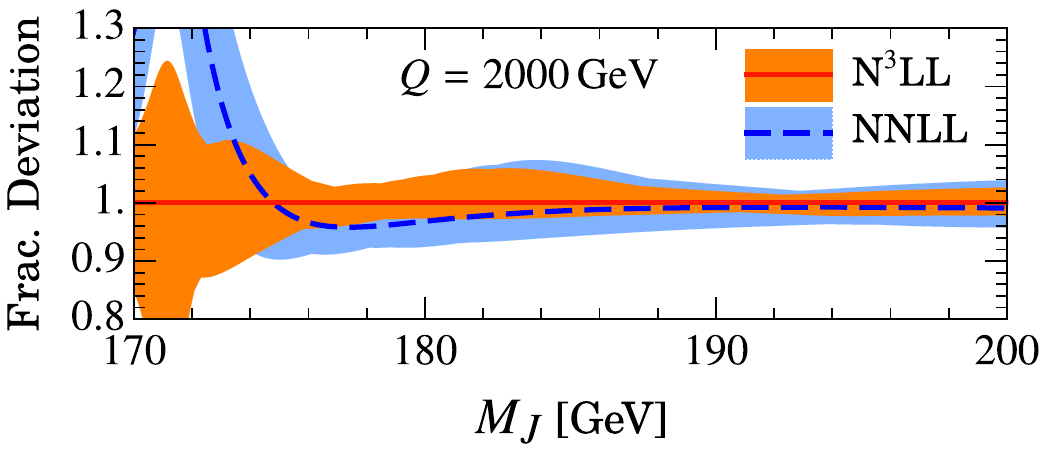}
\vspace{-0.3cm}
\end{center}
\caption{\label{fig:profile-Pole-mu0sqrt2} Perturbative convergence and uncertainty bands for self-normalized cross sections at $Q=700\,{\rm GeV}$ (upper panels) and $Q=2000\,{\rm GeV}$ (lower panels) in the pole mass scheme and without gap subtractions. The two smaller panels show the same results at the two highest orders, but as a fractional deviation from the central N$^3$LL result.}
\end{figure}

In Fig.~\ref{fig:profile-Pole-mu0sqrt2} we show for comparison the analogous results for cross sections in which the pole scheme for the top quark mass is employed without gap subtractions. We will return and discuss this figure in more detail below.

\begin{figure}[t!]
\includegraphics[width = 0.48\textwidth]{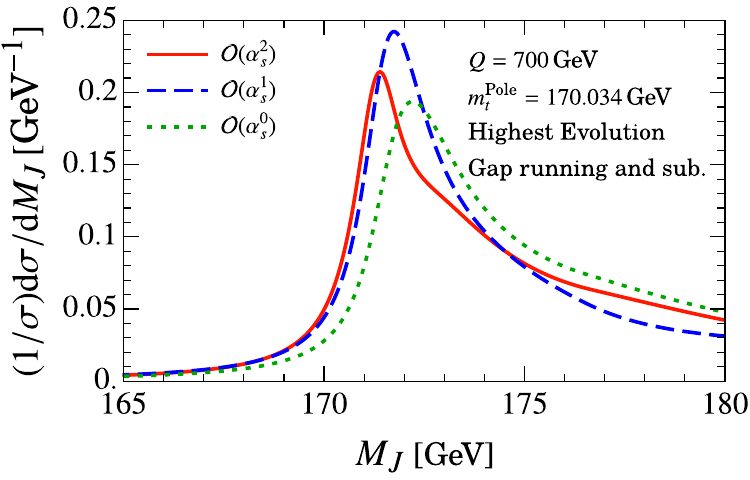}
\includegraphics[width = 0.48\textwidth]{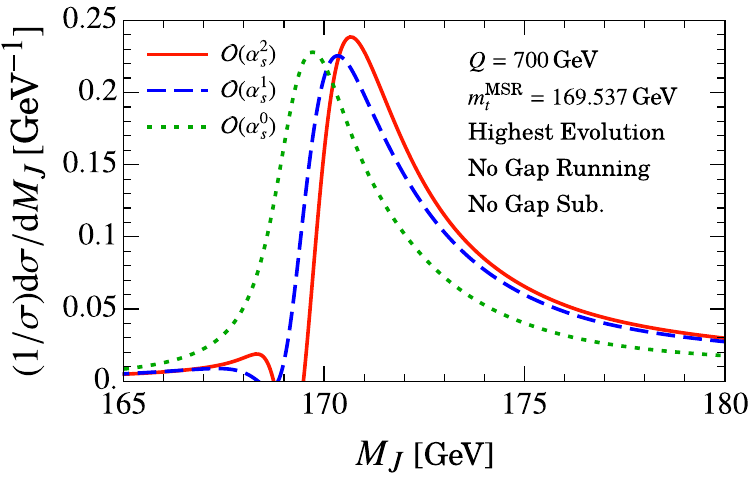}
\includegraphics[width = 0.48\textwidth]{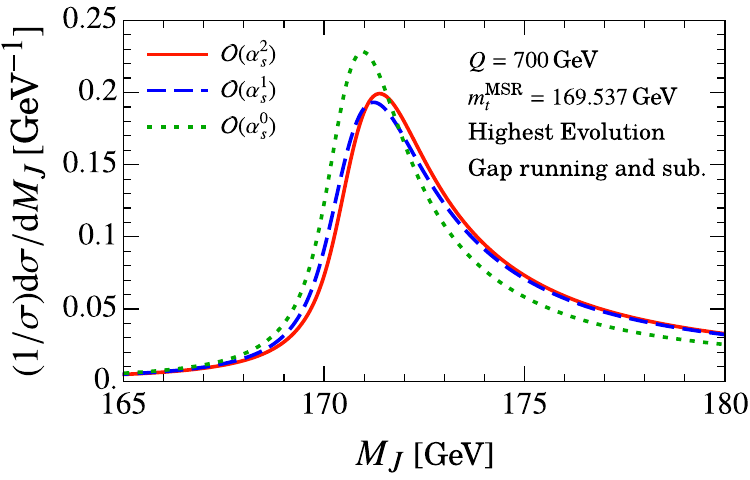}
\caption{\label{fig:renormalon-jet-function}
Analysis of the pole mass and soft function renormalon's impact on the convergence of perturbation theory at ${\cal O}(\alpha_s^0)$, ${\cal O}(\alpha_s)$, and ${\cal O}(\alpha_s^2)$. The top panel uses the jet function in the pole mass scheme, but includes gap subtractions for the soft function. The middle panel does not include gap subtractions, but utilizes the jet function in the MSR mass scheme. The bottom panel employs the MSR mass scheme and gap subtractions, and thus removes the leading renormalons in both the jet and soft functions.}
\end{figure}

It is instructive to first examine the importance and interplay of the ${\cal O}(\Lambda_{\rm QCD})$ renormalons contained in the perturbative fixed-order series of the bHQET jet and partonic soft functions.
To illustrate the impact of the pole mass renormalon in the bHQET jet function, we display in the upper panel of Fig.~\ref{fig:renormalon-jet-function} the 2-jettiness cross section in the {\it pole mass scheme} for $Q=700$\,GeV using the default profile functions, consistently expanding all fixed-order matrix elements entering the factorization theorem (i.e.\ the bHQET jet function, the soft function and the hard function) to ${\cal O}(\alpha_s^0)$ (dotted green), ${\cal O}(\alpha_s)$ (dashed blue), and ${\cal O}(\alpha_s^2)$ (solid red) and consistently using gap subtractions. Thus, renormalon subtractions associated with the pole mass renormalon are not included, while the soft function renormalon is still removed systematically.
Since the renormalization group evolution, which predominantly effects the normalization, does not contain any renormalon effects, we adopt the highest order N$^3$LL anomalous dimensions for all renormalization-group resummation factors, so that the behavior of the three curves is focused on the pole mass renormalon in the unsubtracted jet function. The curves clearly
exhibit the well-known pole mass renormalon problem which causes the peak position to systematically shift towards smaller jet masses with increasing order. At the level of the two-loop bHQET jet function itself, this behavior was discussed in~\cite{Jain:2008gb}. In a fit to data this behavior would correspond to a pole mass value that systematically increases with the perturbative order. This is the known behavior of the perturbative series for the pole mass in terms of a short distance mass~\cite{Hoang:2017btd,Beneke:2016cbu,Hoang:2020iah}. Furthermore, the curves show some instabilities in its shape, in particular in the form of the distribution at and above the peak.

In order to illustrate the impact of the soft function renormalon we display in the middle panel of Fig.~\ref{fig:renormalon-jet-function} the 2-jettiness cross section {\it without gap subtractions}, again for $Q=700$\,GeV using the default profile functions and consistently expanding all fixed-order matrix elements entering the factorization theorem, but this time using the MSR top quark mass scheme (and the standard $\overline{\rm MS}$ mass in the hard function). Here subtractions associated to the soft function renormalon are not included, while the pole mass renormalon is removed systematically. As in the upper panel, we adopt the highest order N$^3$LL anomalous dimensions for all renormalization-group resummation factors, so that the three curves focus on the behavior due to the soft function renormalon. We see that the soft function renormalon causes the peak position to systematically shift towards larger jet masses with increasing order. In a fit to data this behavior would correspond to an MSR mass value that systematically decreases with the perturbative order. Furthermore, the curves at ${\cal O}(\alpha_s)$ and ${\cal O}(\alpha_s^2)$ show considerable shape instabilities in the region below the peak, where the cross section can even become negative. We note that the impact of the soft function renormalon increases with the c.m.\ energy $Q$. This dependence on $Q$ arises from the boost factor $\vel=Q/m_t$ appearing in the convolution integral shown in \eq{fact}, which is also manifest in Eq.~(\ref{eq:totalsub}).

Finally, in the lower panel of Fig.~\ref{fig:renormalon-jet-function} we have displayed the corresponding three curves once again, but systematically accounting for the subtractions associated to both the pole mass and soft function renormalons, by using the MSR top quark mass scheme and gap subtractions, respectively. We now observe very good convergence of the peak position and, furthermore, no instabilities in the shape of the distribution are visible.

The upper and middle panels of Fig.~\ref{fig:renormalon-jet-function} also nicely illustrate the presence of a partial cancellation of the jet and soft function renormalon effects since they have opposite signs. In the combined order-by-order cross sections both corrections thus partially cancel when the pole mass scheme is employed and no gap subtraction is applied for the soft function. Even though this partical cancellation arises between two independent physical effects and should therefore be considered as accidental from a principle point of view, it does undeniably take place in the physical regions of c.m.\ energies where high-precision extractions of the top mass can be carried out. One may therefore ask the question whether this cancellation may in principle allow for a pole mass determination where the impact of the pole mass renormalon could be tamed or even avoided altogether. At this point we would like to remind the reader that for a top mass determination from data (or MC pseudo data) simultaneous fits of the peak region 2-jettiness distribution for several $Q$ values are needed to disentangle the dependence on the top quark mass and the shape function parameters. So there is a strong degeneracy concerning the dependence on the top quark mass and the shape function parameters, and in particular its first moment $\Omega_1$. Given that there are strong cancellations between corrections affecting both of these dependences, it can be expected that they degrade the overall precision of such an analysis. Furthermore, since the amount of mutual cancellation between the pole mass and soft function renormalons is $Q$-dependent, it is expected that such fits for theoretical predictions without any renormalon subtractions will exhibit larger theoretical uncertainties compared to those where the pole mass and soft function renormalons are separately and independently subtracted. Such an extensive analysis is, however, beyond the sope of this work.

Furthermore, it should also be pointed out that the shape function appearing in Eq.~\eqref{eq:fact} is universal and appears in the same form also in the factorization theorem for the $e^+e^-$ thrust distribution below the top pair threshold, where precise information on its parameters can be extracted from available $e^+e^-$ data~\cite{Abbate:2010xh} at significantly smaller values of $Q$. The thrust distribution for massless quark production is sensitive to the same soft function renormalon, but does not have any top mass dependence. Thus if information on the renormalon-free shape function parameters obtained from $e^+e^-$ thrust data are systematically accounted for, it is unavoidable that renormalon effects must be properly handled for top quark mass determinations from the \mbox{2-jettiness} distribution.

Finally, it is instructive to also have a closer look at the 2-jettiness cross section without any renormalon subtraction. In Fig.~\ref{fig:profile-Pole-mu0sqrt2} the 2-jettiness differential cross section at $Q=700$\,GeV (upper panels) and $Q=2000$\,GeV (lower panels) in the pole mass scheme and without gap subtractions are shown for the default profile functions. The results are normalized to the $M_J$ interval displayed in the respective panels at NLL (green dotted line), NNLL (blue dash-dotted line) and N$^3$LL (red solid line). We also display uncertainty bands with the corresponding colors at each of the three orders. These bands are again derived by determining the upper and lower value of the distributions (for each $M_J$ value) obtained by considering $500$ profile functions generated randomly within the profile function parameter ranges. Apart from the fact that neither the pole mass nor the soft function renormalons are subtracted, the setup used for all curves and uncertainty bands in Fig.~\ref{fig:profile-Pole-mu0sqrt2} is precisely the same as the one used for Fig.~\ref{fig:profile-MSR-mu0sqrt2}.
\begin{figure}[t!]
\begin{center}
\hspace{-0.15cm}\includegraphics[width = 0.465\textwidth]{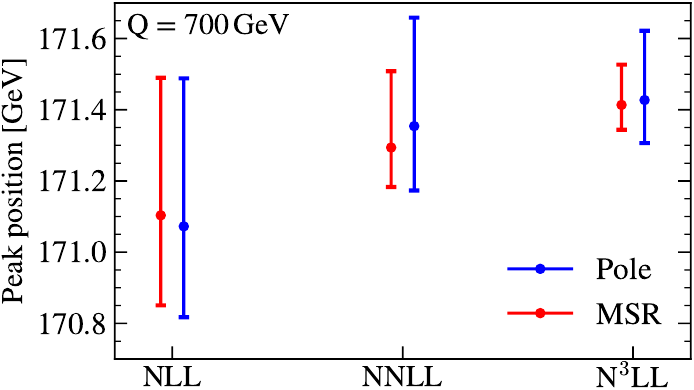}\\\vspace*{0.4cm}
\hspace{0.15cm}\includegraphics[width = 0.45\textwidth]{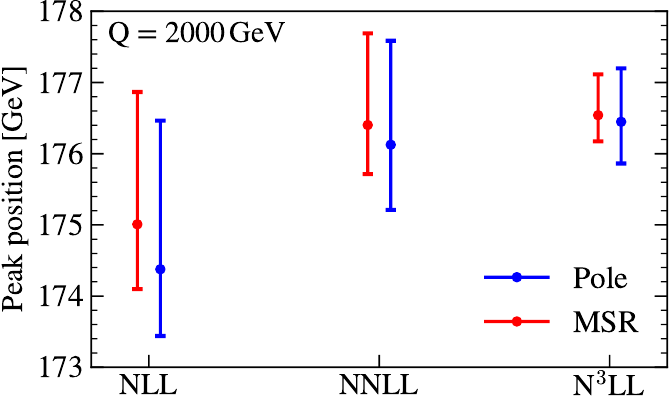}
\end{center}
\caption{\label{fig:peaks} Peak positions at $Q=700\,$GeV (upper panel) and $Q=2000\,$GeV (lower panel) for cross sections in the MSR (red) and pole schemes (blue)
at NLL, NNLL and N$^3$LL accuracy. The error bars are obtained from a flat random scan over $500$ parameters and the central value corresponds to the default profile.}
\end{figure}
\begin{table}[t!]
\begin{tabular}{c c| c c c}
mass & $Q$ & \multicolumn{3}{c}{Peak Positions [GeV]} \\
scheme & [GeV] & NLL & NNLL & N$^3$LL \\
\hline
\multirow{2}{*}{MSR}
& $700$ & $171.104^{+0.386}_{-0.253}$ & $171.294^{+0.214}_{-0.111}$ & $171.414^{+0.113}_{-0.070}$ \\
& $2000$ & $175.008^{+1.858}_{-0.910}$ & $176.403^{+1.287}_{-0.690}$ & $176.541^{+0.574}_{-0.367}$ \\
\multirow{2}{*}{Pole}
& $700$ & $171.073^{+0.416}_{-0.255}$ & $171.354^{+0.305}_{-0.181}$ & $171.427^{+0.195}_{-0.121}$ \\
& $2000$ & $174.377^{+2.087}_{-0.938}$ & $176.126^{+1.461}_{-0.915}$ & $176.448^{+0.750}_{-0.587}$
\end{tabular}
\caption{\label{tab:peaks} Peak positions at different perturbative orders using the MSR and pole mass schemes, as shown in Fig.~\ref{fig:peaks}.}
\end{table}
We see that the perturbative behavior concerning the convergence and the perturbative uncertainties is also good even without any renormalon subtraction. This underlines the partial cancellation of the jet and soft function renormalons. However, a closer inspection shows that the perturbative uncertainty bands are narrower when the subtraction of all renormalons is taken care of systematically. This is visible in the fractional deviation plots, where the N$^3$LL renormalon-subtracted predictions in Fig.~\ref{fig:profile-MSR-mu0sqrt2} exhibit an average uncertainty of $\pm 3.8\%$ at $Q=700$\,GeV for $M_J\ge 171$\,GeV compared to $\pm 5.5\%$ for the predictions without any renormalon subtraction in Fig.~\ref{fig:profile-Pole-mu0sqrt2}.
In contrast, for $Q=2000\,{\rm GeV}$ Fig.~\ref{fig:profile-MSR-mu0sqrt2} has an average uncertainty of $\pm 2.4\%$ for $M_J\ge 175\,{\rm GeV}$ compared to $\pm 2.9\%$ for the predictions without any renormalon subtraction in Fig.~\ref{fig:profile-Pole-mu0sqrt2}.

An interesting aspect of our definition for the jet mass variable $M_J$ [\,defined using 2-jettiness in \eq{MJdef}\,], is that it is normalized in a way such that it can be seen as a direct measure for the top quark mass. Therefore, the behavior of the peak position for the $M_J$ distribution allows us to draw conclusions on the size of the perturbative uncertainties of a top mass determination from the peak position. To avoid outliers we discard the two highest and two lowest points in the scan so as to better represent the bulk of the points. In Fig.~\ref{fig:peaks} we show the peak positions of the curves at $Q=700$\,GeV and $Q=2000$\,GeV for the default profile functions and their perturbative uncertainties, estimated from the $500$ random profile functions. Results are shown in the MSR mass scheme with gap subtraction (MSR, red) and in the pole mass scheme without gap subractions (Pole, blue) at NLL, NNLL and N$^3$LL. The central values, shown by dots, correspond to the default profile scales, so the perturbative uncertainties are asymmetric. The corresponding numbers are also given in Tab.~\ref{tab:peaks}. The results show that the perturbative uncertainty is systematically smaller when the top quark mass and soft function renormalons are subtracted. For $Q=700$\,GeV, where we have the highest top quark mass sensitivity, using renormalon subtractions leads to an uncertainty in the peak location of around $\pm 85$\,MeV at N$^3$LL order. Without renormalon subtractions the uncertainty at this order increases to around $\pm 150$\,MeV, which is almost factor of two larger. For $Q=2000$\,GeV the uncertainties are larger for the analysis without renormalon subtraction as well (around $\pm 450$\,MeV with renormalon subtraction compared to around $\pm 650$\,MeV without renormalon subtractions, both at N$^3$LL order). Here the difference is less pronounced because the overall top mass quark sensitivity decreases for larger $Q$ values and the overall uncertainties increase. The analogous behavior is also visible at lower orders.
Our results indicate that the MSR mass may be extracted with an uncertainty of well below $100$\,MeV, while the pole mass uncertainty is at the level of $150$\,MeV. Interestingly, this is about the size of the top quark pole mass renormalon ambiguity of $166$\,MeV that was estimated recently in Ref.~\cite{Hoang:2017btd}
for the case of massless charm and bottom quarks (which is the approximation we use in our analysis). Note that in an earlier analysis in Ref.~\cite{Beneke:2016cbu} the top quark pole mass renormalon ambiguity was estimated as the smaller value of $67$\,MeV (also for massless charm and bottom quarks).

The results we have obtained in this simple analysis of the peak positions do --- taken by themselves --- not contradict the view that the top quark pole mass can be extracted from the 2-jettiness cross section with perturbative uncertainties below the pole mass renormalon ambiguity, but they also show that at least at N$^3$LL order the precision is not (yet) sufficient to achieve that goal and that higher-order corrections beyond this order would be mandatory to get there. On the other hand, the results also
support the view that, even though the 2-jettiness cross section exhibits a cancellation between the pole mass and soft function renormalons, the pole mass can still not be extracted with a precision below its renormalon ambiguity. In any case, using renormalon subtractions, and in particular the MSR mass scheme, will yield substantially higher precision and smaller perturbative uncertainties.

At this point we would like to again mention, that
in mass determinations from data (or MC pseudo data) simultaneous fits of the peak region 2-jettiness distribution for several $Q$ values are needed to disentangle the dependence on the top quark mass and the shape function parameters, and that the whole distribution in the peak region (rather than just the peak position) would enter such fits.
As mentioned before, however, this kind of study requires that also off-shell and $m_t/Q$ power-suppressed contributions are included, as their effects can be non-negligible depending on how the cross section is normalized.

The dominant such QCD corrections to the factorization theorem in the bHQET region come from two sources, mass power corrections appearing as higher order terms in \eq{tauM}, and corrections to the perturbative singular structures. (Additional non-singular kinematic power corrections are very small at one-loop and hence irrelevant.) The former are universal at any order in $\alpha_s$ and shift the distribution to the right by ${\cal O}(m_t^4/Q^4)$ but are trivial to incorporate.
The latter are known analytically to ${\cal O}(\alpha_s)$ is QCD~\cite{Lepenik:2019jjk}:
at tree-level one gets a modification of the coefficient of the delta function, while at $\mathcal{O}(\alpha_s)$ also the plus distribution coefficient is affected.
One can include these mass corrections by a suitable modification of the hard and jet functions (see e.g.\ Ref.~\cite{Bris:2020uyb} for more details).
These mass power corrections decrease the cross sections in Fig.~\ref{fig:renorm-corr} by $5\%$ (beyond NLL) everywhere except for the region to the left of the peak where the effects are smaller.
However if the cross section is normalized the effect of these power corrections drops below a percent becoming negligible in all relevant regions.
It is reasonable to believe that these power corrections will be of similar form and size once ${\cal O}(\alpha_s^2)$ are added.
A complete analysis that accounts for these effects will require the inclusion of the ${\cal O}(\alpha_s^2)$ correction to the primary massive quark SCET jet function that was computed recently in Ref.~\cite{Hoang:2019fze} and shall be carried out in future work.

\section{Conclusions}
\label{sec:conclusion}

In this article we have presented results for the \mbox{2-jettiness} differential distribution for boosted tops produced in $e^+e^-$ collisions in the peak region, accounting for the resummation of large QCD logarithms at next-to-next-to-next-to-leading logarithmic (N$^3$LL) order and fixed-order corrections to the hard, soft and jet function matrix elements at next-to-next-to-leading order [\,${\cal O}(\alpha_s^2)$\,], calculated in the framework of soft-collinear effective theory and boosted heavy quark effective theory. We have systematically removed the ${\cal O}(\Lambda_{\rm QCD})$ renormalons contained in the soft and jet functions, by using a gap subtraction as well as the MSR mass, and have provided a numerical analysis indicating that the perturbative uncertainties of a determination of the top quark MSR mass from the N$^3$LL\,+\,$\mathcal{O}(\alpha_s^2)$ prediction at a c.m.\ energy of $Q=700$\,GeV are well below the level of $100$\,MeV. For future reference all theoretical formulae have been given explicitly in several appendices.

An interesting aspect of the 2-jettiness distribution is that the soft- and jet-function renormalons partially cancel each other for center-of-mass energies above $700$\,GeV where the boosted top quark approximation is valid and precise top mass determinations can be carried out. This cancellation arises because the soft-function and pole-mass renormalons enter with different signs. While these two renormalons represent two physically independent infrared sensitivities, the cancellation allows for rather stable and convergent predictions in the pole mass scheme, if at the same time also the soft function renormalon is left unsubtracted. However, the resulting perturbative uncertainties are still systematically larger compared to the predictions where both renormalons are independently removed.

The analysis done here based on boosted heavy quark effective theory neglects subleading collinear off-shell corrections, which have been determined recently at ${\cal O}(\alpha_s^2)$ in Ref.~\cite{Hoang:2019fze} and shall be accounted for in future work.

\begin{acknowledgments}
This work was supported in part by FWF Austrian Science Fund under the Project No.~P28535-N27, the Spanish MINECO Ram\'on
y Cajal program (RYC-2014-16022), the Office of Nuclear Physics of the U.S. Department of Energy under the Grant No. DE-SCD011090,
the MECD grants FPA2016-78645-P and PID2019-105439GB-C22, the IFT Centro de Excelencia Severo Ochoa Program under
Grant SEV-2012-0249, the EU STRONG-2020 project under the program H2020-INFRAIA-2018-1, grant agreement no. 824093
and the COST Action CA16201 PARTICLEFACE. I.S. was also supported by the Simons Foundation through the Investigator grant 327942.
We acknowledge partial support by the FWF Austrian Science Fund under the Doctoral Program “Particles and Interactions” No. W1252-N27.
A.P. is a member of the Lancaster-Manchester-Sheffield Consortium for Fundamental Physics, which is supported by the UK Science and Technology Facilities Council (STFC) under grant number ST/T001038/1.
VM thanks the University of Vienna and MIT for hospitality while parts of this work were completed.
BB is partially supported by the Government of the Republic of Trinidad and Tobago.
BB also thanks the University of Vienna and DESY for hospitality while parts of this work were completed.
Figures are made using the \Mathematica package \PLHot~\cite{plhot}.
\end{acknowledgments}

\appendix
\section{Formulae}
\label{app:formulae}

The factorization theorem presented in \eq{fact} is expressed in momentum space where the jet and soft functions, along with the evolution factors, are distributions involving series in plus and delta functions. We find it convenient to combine the ingredients in position space where the convolutions become simple products.
Below our notation and definitions with variable mass dimension follow Ref.~\cite{Fleming:2007xt}. For earlier work on the associated resummation formulae see Refs.~\cite{Korchemsky:1993uz,Balzereit:1998yf,Neubert:2004dd}.
For a function ${\cal F}(q,\mu)$ that depends on a momentum-space variable $q$ with mass dimensions $j_{\cal F}$ and the renormalization scale $\mu$, the Fourier transform in position space is defined as
\begin{align}\label{eq:FT}
\tilde {\cal F}(x,\mu) = \int_{-\infty}^{\infty} \df q \: e^{-i q x} {\cal F}(q, \mu) \, ,
\end{align}
where $x$ has mass dimensions $-j_{\cal F}$. The position-space anomalous dimension permits writing the corresponding RG relation as a local equality:
\begin{equation}
\mu \frac{\df }{\df \mu} \widetilde{\cal F} (x,\mu) =
\widetilde \gamma_{\cal F}(x, \mu)\, \widetilde{{\cal F}}(x, \mu)\,,
\end{equation}
such that RG-evolved position-space soft and jet functions can be expressed as a regular product:
\begin{align}
{\cal F}(q, \mu) &= \int\! \df q^\prime\: U_{\cal F}(q - q^\prime, \mu, \mu_0) \, {\cal F}(q^\prime, \mu_0) \\
&= \int\! \frac{\df x}{2 \pi} \:e^{i q x}\,\tilde U_{\cal F}(x, \mu, \mu_0) \, \tilde {\cal F}(x, \mu_0) \, .\nn
\end{align}

The factorization function $H_Q$ is not a distribution, but a simple function of the center of mass energy $Q$, with mass dimension $j_H=1$ in our convention.
One can treat factorization, and position-space jet and soft functions on the same footing simply using ${\cal Q}=\{Q,1/(ie^{\gamma_E} x)\}$, \mbox{$F=\{H_Q,\tilde {\cal F}\}$} and $\gamma_F=\{\gamma_H, \tilde{\gamma}_{\cal F}\}$ [\,we describe the evolution of the bHQET current in \eq{bHQET-run}\,]. In this way, we express the RGE and evolution from $\mu_0$ to $\mu$ as
\begin{align}\label{eq:posRGE2}
\mu \frac{\df}{\df\mu} F({\cal Q},\mu) &=
\Bigl[\Gamma_{\!F}[\alpha_s] \log\Bigl(\frac{\mu}{{\cal Q}}\Bigr) + \gamma_F[\alpha_s] \Bigr] F(\mu, \mu_0) \, , \nn
\\
F({\cal Q},\mu) &= U_{\!F}(\mu,\mu_0; {\cal Q}) F(\mu_0, {\cal Q})
\\
&\equiv e^{K_{\!F}(\mu, \mu_0) } \, \Bigl(\frac{\mu_0^{j_F}}{{\cal Q}}\Bigr)^{\!\omega_F(\mu,{\cal Q})} F({\cal Q},\mu_0) \, ,\nn
\end{align}
where $\Gamma_{\!F}[\alpha_s]$ resums double logarithms and is proportional to the universal cusp anomalous dimension $\Gamma^{\rm cusp}[\alpha_s]$, and $\gamma_F[\alpha_s]$ is the noncusp anomalous dimension.
The evolution kernels $K_{\! F}$ and $\omega_{F}$ are defined as [$\alpha_0\equiv \alpha_s(\mu_0)$, $\alpha_\mu\equiv \alpha_s(\mu)$]
\begin{align}
\label{eq:evolKernels}
K_{\! F}(\mu, \mu_0) &= K^{\Gamma}_{\! F}(\mu, \mu_0) + K^{\gamma}_{\! F}(\mu, \mu_0)\,, \\
K^{\Gamma}_{\! F} (\mu, \mu_0) &= {j_{F}}\!\int_{\alpha_0}^{\alpha_\mu} \frac{\df \alpha}{\beta (\alpha)} \Gamma_{\! F}[\alpha] \!\int_{\alpha_0}^{\alpha} \frac{\df \alpha^\prime}{\beta (\alpha^\prime)} \,,\nn \\
K^{\gamma}_{\! F} (\mu, \mu_0) &= \!\int_{\alpha_0}^{\alpha_\mu} \frac{\df \alpha}{\beta (\alpha)} \gamma_{ F}[\alpha] \, , \nn\\
\omega_{F}(\mu, \mu_0) &= \!\int_{\alpha_0}^{\alpha_\mu} \frac{\df \alpha}{\beta (\alpha)} \Gamma_{\! F}[\alpha] \, . \nn
\end{align}
The results of the evolution kernels at N$^3$LL are given by
\begin{align}
K^{\Gamma}_{\! F}(\mu,\mu_0) &={j_{F}} K\big(\Gamma_{\! F}, \mu, \mu_0\big) \, , \\ K^{\gamma}_{F}(\mu,\mu_0) &= \eta (\gamma_{ F}, \mu, \mu_0) \, ,\nn \\
\omega_{F}(\mu,\mu_0) &= \eta(\Gamma_{\! F}, \mu, \mu_0) \, , \nn
\end{align}
where
\begin{widetext}
\begin{align} \label{eq:w}
\eta(\Gamma,\mu,\mu_0) &= -\frac{\Gamma_{\!0}}{2\beta_0}\biggl\{\log r
+\frac{\alpha_0}{4\pi}\biggl(\frac{\Gamma_{\!1}}{\Gamma_{\!0}}
-\frac{\beta_1}{\beta_0}\biggr)(r-1)
+\frac{1}{2}
\biggl(\frac{\alpha_0}{4\pi}\biggr)^{\!\!2}\biggl(\frac{\beta_1^2}{\beta_0^2}
-\frac{\beta_2}{\beta_0}+\frac{\Gamma_{\!2}}{\Gamma_{\!0}}
-\frac{\Gamma_{\!1}\beta_1}{\Gamma_{\!0}\beta_0}\biggr)(r^2-1) \\
& \qquad+\frac{1}{3}
\biggl(\frac{\alpha_0}{4\pi}\biggr)^{\!\!3}
\biggl[\frac{\Gamma_{\!3}}{\Gamma_{\!0}}-\frac{\beta_3}{\beta_0}
+\frac{\Gamma_{\!1}}{\Gamma_{\!0}}\biggl(\frac{\beta_1^2}{\beta_0^2}-\frac{\beta_2}{\beta_0}\biggr)
-\frac{\beta_1}{\beta_0}\biggl(\frac{\beta_1^2}{\beta_0^2}-
2\,\frac{\beta_2}{\beta_0}+\frac{\Gamma_{\!2}}{\Gamma_{\!0}}\biggr) \biggr](r^3-1)\biggr\},\nn
\end{align}
and
\begin{align} \label{eq:K}
&K(\Gamma,\mu,\mu_0)
=\frac{\Gamma_{\!0}}{4\beta_0^2}\Biggl\{\frac{4\pi}{r \alpha_0}(r\log r+1-r) +
\biggl(\frac{\Gamma_{\!1}}{\Gamma_{\!0}}-\frac{\beta_1}{\beta_0}\biggr)(r-1-\log r)
-\frac{\beta_1}{2\beta_0}\log^2 r \\
&\qquad +\frac{\alpha_0}{4\pi}\biggl[
\biggl(\frac{\Gamma_{\!1}\beta_1}{\Gamma_{\!0}\beta_0}-\frac{\beta_1^2}{\beta_0^2}\biggr)
(r-1-r \log r)
-B_2 \log r
+\biggl( \frac{\Gamma_{\!2}}{\Gamma_{\!0}}-\frac{\Gamma_{\!1}\beta_1}{\Gamma_{\!0}\beta_0} +
B_2 \biggr)\frac{r^2-1}{2}
+\biggl(\frac{\Gamma_{\!1}\beta_1}{\Gamma_{\!0}\beta_0}-\frac{\Gamma_{\!2}}{\Gamma_{\!0}}\biggr)
(r-1)\biggr]
\nn\\
& \qquad
+\biggl(\frac{\alpha_0}{4\pi}\biggr)^{\!\!2}
\biggl[ \biggl(B_2\biggl(\frac{\Gamma_{\!1}}{\Gamma_{\!0}}-\frac{\beta_1}{\beta_0} \biggr)
+\frac{B_3}{2}\biggr)\frac{r^2\!-\!1}{2}
+ \biggl(\frac{\Gamma_{\!3}}{\Gamma_{\!0}} -
\frac{\Gamma_{\!2}\beta_1}{\Gamma_{\!0}\beta_0}
+\frac{B_2\Gamma_{\!1}}{\Gamma_0}+B_3\biggr) \Bigl(\frac{r^3-1}{3}-\frac{r^2-1}{2}\Bigr)
\nn\\
&\qquad \qquad\qquad
-\frac{\beta_1}{2\beta_0}
\biggl(\frac{\Gamma_{\!2}}{\Gamma_{\!0}}-\frac{\Gamma_{\!1}\beta_1}{\Gamma_{\!0}\beta_0}+B_2\biggr)
\Bigl(r^2\log r-\frac{r^2-1}{2}\Bigr) -\frac{B_3}{2}\log r
-B_2\biggl(\frac{\Gamma_{\!1}}{\Gamma_{\!0}}-\frac{\beta_1}{\beta_0} \Bigr)(r-1)
\biggr]
\Biggr\},\nn
\end{align}
\end{widetext}
where $r=\alpha_\mu/\alpha_0$ depends on the 4-loop running coupling, and the $B_i$ coefficients take the following values: $B_2=\beta_1^2/\beta_0^2-\beta_2/\beta_0$ and $B_3=2\beta_1\beta_2/\beta_0^2-\beta_1^3/\beta_0^3-\beta_3/\beta_0$. The series expansions of the QCD beta function and the cusp and noncusp pieces of a generic SCET anomalous dimensions are written as
\begin{equation}
\!\!\!\!\beta[\alpha_s]=-2\alpha_s\!\!\sum_{n=0}^{\infty}\beta_n\Bigl(\frac{\alpha_s}{4\pi}\Bigr)^{\!n+1}\!,\,\,\,
\Gamma[\alpha_s]=\!\sum_{n=0}^{\infty}\Gamma_{\!n}\Bigl(\frac{\alpha_s}{4\pi}\Bigr)^{\! n+1}\!,
\end{equation}
where here $\Gamma$'s either stand for $\Gamma_{\! F}$, $\Gamma^{\rm cusp}$ or
$\gamma_F$. The numerical expressions for the universal cusp anomalous dimension coefficients for $5$ and $6$ flavors are given by~\cite{Korchemsky:1987wg,Moch:2004pa,Henn:2019swt}
\begin{align}
\bigl\{\Gamma_i^{{\rm cusp}} \bigr\}^{(5)}_{0\leq i\leq3}&=\{4 , \, 27.633,\, 179.406 ,\, 141.254\} \, ,\\
\bigl\{\Gamma_i^{{\rm cusp}} \bigr\}^{(6)}_{0\leq i\leq3}&=\{4 , \, 23.188 ,\, 35.497, \,-2581.527 \} \, .\nn
\end{align}
where the 4-loop cusp anomalous dimension is obtained from recent work in Refs.~\cite{Moch:2018wjh,Henn:2019swt}. Note that in our convention we do not include a factor of $C_F$ in the definition of $\Gamma^{\rm cusp}[\alpha_s]$, but we do include this factor for the
$\Gamma_{\!\!F}[\alpha_s]$'s for various functions given below in App.~\ref{app:anomDim}.
Finally, we quote the numerical results for the QCD beta function up to 4 loops~\cite{Tarasov:1980au, Larin:1993tp,
vanRitbergen:1997va,Czakon:2004bu}:
\begin{align}
\bigl\{ \beta_i^{(5)} \bigr\}_{0\leq i\leq3}
&=\{23/3\,, 116/3\,,180.907\,,4826.16\} ,\nn \\
\bigl\{ \beta^{(6)}_i \bigr\}_{0\leq i\leq3}&=\{ 7\,, 26 \,, -32.5\,,2472.28\} .
\end{align}
\section{Anomalous Dimensions}\label{app:anomDim}

The RGE of various functions appearing in \eq{fact} are as follows:
\begin{align}\label{eq:RGEall}
\mu\frac{\df}{\df \mu}\! \log \bigl[H_{Q}^{(6)}(Q, \mu) \bigr] &\!=\!
\Gamma_{\!H_Q}[\alpha_s]\log \Bigl(\frac{\mu}{Q}\Bigr)\! +\! \gamma_{H_Q}[\alpha_s]
\,, \\
\mu\frac{\df}{\df \mu} \!\log \bigl[{\cal J}_{\!v}^{(5)}(\vel,\mu) \bigr]
&\!=\!
\Gamma_{\!v}[\alpha_s] \log\Bigl(\frac{1}{\vel}\Bigr) + \gamma_{v}[\alpha_s]\,, \nn
\\
&\equiv \gamma_{v}(\vel)\,,\nn\\
\mu\frac{\df}{\df \mu} \!\log \bigl[\tilde J^{(5)}_{\!B,\tau_2}\!(x, \mu) \bigr]
&\!=\!
\Gamma_{\!\!J_{\!B}^{\tau}}\![\alpha_s] \log(ie^{\gamma_E}x \mu)
\!+\!
\gamma_{\!J_{\!B}^{\tau}}\![\alpha_s]
\,,\nn \\
&\equiv \tilde \gamma_{J_{\!B}^{\tau}}(x,\mu)
\,,\nn \\
\mu\frac{\df}{\df \mu}\!\log \bigl[\tilde S^{(5)}_{\!\tau_2}(y, \mu) \bigr]
&\!=\!
\Gamma_{\!S_{\tau}}[\alpha_s]
\log(ie^{\gamma_E}y \mu) +
\gamma_{S_{\tau}}[\alpha_s] \, ,
\nn\\
&\equiv \tilde \gamma_{S_{\tau}}\!(x,\mu)\,,\nn
\end{align}
where $\tilde J_{B,\tau_2}$ and $\tilde S_{\tau_2}$ are the Fourier transforms of the thrust, unsubtracted stable bHQET jet and soft functions defined in \eqs{JBtau2stable}{Spositionspace}. Note that we have $\gamma_{H_Q}=2\gamma_H$, $\gamma_{\!J_{\!B}^{\tau}} = 2\gamma_B$ and $\gamma_{S_{\tau}}=2\gamma_S$ with $\gamma_{H,S}$ given as in Ref.~\cite{Abbate:2010xh} and $\gamma_{B}$ defined in Ref.~\cite{Fleming:2007xt}. In \eq{fact}, instead of running $H_m$ we RG evolve the squared matrix element of the bHQET current, ${\cal J}_{v}$, defined as
\begin{align}
{\cal J}^{(5)}_v = |\langle{\cal J}^{(5)}_{\rm bHQET} \rangle |^2 \, .
\end{align}
The RG evolution of ${\cal J}^{(5)}_v$, using the results in \eq{RGEall} for the evolution factor $U_v$ shown in \eq{Hevol}, reads:
\begin{equation}\label{eq:bHQET-run}
U_v (\vel, \mu_m, \mu) =e^{ K_{v}^{\gamma(5)}(\mu_m, \mu)} \vel^{-\omega^{(5)}_{v}\!(\mu_m, \mu)}
\,,
\end{equation}
where, keeping in line with our notation, we emphasize that $\mu_m$ is the final scale up to which ${\cal J}^{(5)}_v$ is RG evolved.
The bHQET current ${\cal J}^{(5)}_{\rm bHQET}$~\cite{Fleming:2007xt,Hoang:2015vua} is given by
\begin{align}
{\cal J}_{\rm bHQET} = \bar h_{v_t} W_n Y_n^\dagger \Gamma_i^{\mu} Y_{\bn} W_{\bn}^\dagger h_{v_\tb} \, ,
\end{align}
where $h_{v_{t,\tb}}$ are the heavy-quark fields describing top and anti-top quarks; $W_{n,\bn}$ are the Wilson lines formed from ultracollinear gluons, such that in position space $W_n^\dagger(x) = {\rm P} \exp\bigl[ig\int_0^\infty {\rm d}s\, \bn \cdot A_n(\bn s + x) \bigr]$; $Y_{n,\bn}$ are similarly defined Wilson lines with ultrasoft gluons, and $\Gamma_v^\mu = \gamma^\mu$ and $\Gamma_a^\mu = \gamma^\mu \gamma_5$. The RG consistency in the bHQET sector implies the following constraint:
\begin{align}\label{eq:v-gamma}
\gamma_v(\vel)
&=\tilde \gamma_{J_{\!B}^{\tau}}(x,\mu) + \tilde \gamma_{S_{\tau}}\!\bigl(\vel x, \mu\bigr) \, ,
\end{align}
with $\gamma_v(\vel)$, $\tilde \gamma_{J_{\!B}^{\tau}}(x,\mu)$, and $\tilde \gamma_{S_{\tau}}\!\bigl(\vel x, \mu\bigr)$ defined above in \eq{RGEall}.
This leads to a cancellation of the $\log(\mu)$ dependence in the anomalous dimension of the bHQET current and implies
\mbox{$\Gamma_{\!v}[\alpha_s] = \Gamma_{\!\!J_{B}^{\tau_2}}[\alpha_s] =-\Gamma_{\!S_{\tau}}[\alpha_s]$}, \mbox{$\gamma_v[\alpha_s]=\gamma_{S_{\tau}}[\alpha_s]+\gamma_{J_{B}^{\tau}}[\alpha_s]$}.
Since the ``cusp'' piece in \eq{v-gamma} is \mbox{$\mu$-independent}, we find that the kernel $K_{v}^\Gamma$ [\,see \eq{evolKernels}\,] does not appear in the RG evolution of the bHQET matrix element in \eq{HevolRG}, but $K_v^\gamma$ and $\omega_v$ do.
With the convention in \eq{RGEall} the cusp and the noncusp pieces have the following values~\cite{vanNeerven:1985xr,Matsuura:1988sm,Catani:1992ua,Vogt:2000ci,Moch:2004pa,Neubert:2004dd,Moch:2005id,Idilbi:2006dg,Becher:2006mr,Fleming:2007xt,Jain:2008gb,Hoang:2015vua}:
\begin{align} \label{eq:anomDimAll}
\! \Gamma_{\!\!J_{B}^{\tau_2}}\![\alpha_s] = \Gamma_{\!v}[\alpha_s] &=\! -\Gamma_{\!S_{\tau}}[\alpha_s]= 4C_F \Gamma^{{\rm cusp}(5)}[\alpha_s] \,, \\
\Gamma_{\!H_Q}[\alpha_s] &= -4 C_F \Gamma^{{\rm cusp}(6)}[\alpha_s]\,,\nn\\
\bigl\{ \gamma^{H_Q}_i\bigr\}_{0\leq i\leq 2} &=\!\{-16 , 32.669, -21.044\} \, , \nn \\
\bigl\{ \gamma^{v}_i\bigr\}_{0\leq i\leq 2} &=\! \{ 32/3,\, 86.619 ,\, 477.753\} \, , \nn \\
\bigl\{ \gamma^{J_{B}^{\tau_2}}_i \bigr\}_{0\leq i\leq 2} & =\! \{32/3 ,-65.803 , -840.284\} \, ,\nn\\
\bigl\{ \gamma^{S_{\tau}}_i \bigr\}_{0\leq i\leq 2} &= \! \{ 0, 152.422 ,1318.037\} \, .\nn
\end{align}
Using the formulae for the evolution factors in \eq{posRGE2}, the RGEs in \eq{RGEall} and the anomalous dimensions in \eq{anomDimAll}, one arrives at the resummed expressions in \eqs{factposUnSub}{HevolRG}. From \eq{RGEall} we see that all dynamical momentum variables have dimensions of energy so that $j_{F} = 1$ for all the evolution functions that we consider here.

\section{Fixed order results}
\label{app:fixedOrder}
We now state the results for the two-loop matrix elements and factorization functions that appear in our analysis. We show the results for the logarithm of the functions since it simplifies the structure of the results.

\subsection{2-loop results}
The SCET hard matching function $H_Q^{(6)}$ with 6-flavor coupling is given by~\cite{Matsuura:1987wt,Matsuura:1988sm,Gehrmann:2005pd,Moch:2005id,Baikov:2009bg,Lee:2010cg}
\begin{align}\label{eq:HQ}
&\log \bigl[H_Q^{(6)}\!(Q, \mu)\bigr] = \frac{\alpha^{(6)}_s(\mu)}{4\pi}
\Bigl[ 9.372 - 16 L_Q - 10.667 L_Q^2 \Bigr] \nn \\
& +
\biggl[ \frac{\alpha^{(6)}_s(\mu)}{4\pi} \biggr]^2
\Bigl[305.454 - 163.879 L_Q -173.835 L_Q^2\nn \\
&\qquad \qquad\qquad -49.778 L_Q^3\Bigr]\,,
\qquad L_Q = \log \Bigl(\frac{\mu}{Q}\Bigr) \, .
\end{align}

For the hard matching function at the top mass scale $H_m^{(6)}$ we state the result with $m_t$ expressed in either pole mass or $\overline{\rm MS}$ scheme~\cite{Fleming:2007xt,Hoang:2015vua}:
\begin{align}\label{eq:HmPole}
&\log \bigl[H^{\rm pole(6)}_m(m_t^{\rm pole}, \vel, \mu)\bigr] = \\
&\frac{\alpha^{(6)}_s(\mu)}{4\pi} \Bigl[ 15.053-2.667 L_m+ 2.667 L_m^2 \Bigr] \nn \\
& +\biggl[ \frac{\alpha^{(6)}_s(\mu)}{4\pi} \biggr]^2\Bigl[152.578 -45.728 L_m+ 26.5699 L_m^2\nn \\
& - 6.222 L_m^3+ \log (\vel) \bigl(-5.531 -5.926 L_m -1.778 L_m^2\bigr)\Bigr]\,, \nn\\
\label{eq:HmMSbar}
&\log \bigl[H^{\MS(6)}_m(\mbar_t, \vel, \mu)\bigr]
= \log \bigl[H^{\rm pole(6)}_m(\mbar_t, \vel, \mu)\bigr] \\
&\quad+\biggl[ \frac{\alpha^{(6)}_s(\mu)}{4\pi} \biggr]^2(56.889\,L_{\mbar}-28.444)\,.\nn
\end{align}
where $L_m = \log[(m_t^{\rm pole})^2/\mu^2]$, $L_{\mbar} = \log (\mbar_t^2/\mu^2)$,
and $\vel$ has been defined in \eq{velDef}.

The 2-loop fixed order result for unsubtracted, stable-top bHQET jet function in position space is given by~\cite{Fleming:2007xt,Jain:2008gb}
\begin{align}\label{eq:bHQETJet}
& \log \bigl[ m_t^2\tilde J^{(5)}_{B,\tau_2} (x, \mu) \bigr]=
\\ &\qquad\frac{\alpha^{(5)}_s(\mu)}{4\pi}
\Bigl[
15.053 + 10.667 \tilde L_B + 10.667 \tilde L_B^2
\Bigr]
\nn\\
&\qquad + \biggl[\frac{\alpha^{(5)}_s(\mu)}{4\pi}\biggr]^2
\Bigl[
310.954 + 165.012 \tilde L_B + 155.465 \tilde L_B^2
\nn \\
&\qquad \qquad \qquad + 54.518 \tilde L_B^3
\Bigr] \, , \qquad \tilde L_B = \log \bigl(ie^{\gamma_E} x \mu \bigr)
\, .
\nn
\end{align}

Likewise, the unsubtracted soft function in position space reads~\cite{Fleming:2007xt,Monni:2011gb,Kelley:2011ng}
\begin{align}\label{eq:Soft}
&\log \bigl[\tilde S^{(5)}_{\tau_2}(y, \mu) \bigr] =
\frac{\alpha^{(5)}_s(\mu)}{4\pi}
\Bigl[
-13.159 - 10.667 \tilde L_S^2
\Bigr]
\\
&\qquad + \biggl[\frac{\alpha^{(5)}_s(\mu)}{4\pi}\biggr]^2
\Bigl[
-81.361-49.357 \tilde L_S -73.687 \tilde L_S^2
\nn \\
&\qquad \qquad \qquad -54.518 \tilde L_S^3
\Bigr] \, , \qquad \tilde L_S = \log \bigl(ie^{\gamma_E} y \mu \bigr)
\, .
\nn
\end{align}
\subsection{Generating fixed order terms}
We now describe a helpful algorithm that allows one to generate the fixed-order expansion of the position-space matrix elements and factorization functions discussed here from the non-logarithmic coefficients and their anomalous dimensions in \eq{anomDimAll}. This applies to the matching coefficient $H_Q^{(6)}$, the bHQET jet and the soft matrix elements, and can be used to reproduce the results stated above in Eqs.~(\ref{eq:HQ}), (\ref{eq:bHQETJet}) and (\ref{eq:Soft}). Note that the algorithm described below must be generalized in an obvious way for $H_m^{(6)}$ to obtain the results in \eqs{HmPole}{HmMSbar}, since its running results from the anomalous dimensions of the bHQET current and the SCET matching coefficient (with different number of dynamical flavors in the running coupling) and includes an additional rapidity logarithm.

Consider a position-space matrix element
or a factorization function $F(\mu,{\cal Q})$ having the generic form
\begin{align}
F(\mu,{\cal Q})
=A\exp
\biggl\{\sum_{m = 1} \!a^{ { F}}_{mn} \!\biggl[\frac{\alpha_s(\mu)}{4\pi}\biggr]^{\!m} \sum_{n = 0}^{m+1}
\log^{n}\!\biggl(\frac{\mu^{j_F}}{\cal Q}\biggr)\!\!
\biggr\} ,
\end{align}
where we have $A=1/m_t^2$ for the thrust position-space bHQET jet function and $A=1$ in the other cases,
see \eq{lnSBcoeffs}.
The constant terms $a^{{F}}_{m0}$ serve as independent data, whereas other coefficients can then be determined by anomalous dimensions and the beta function.
Thus the $a^{{F}}_{m0}$ serve as boundary condition data for the RG differential equations in \eq{posRGE2}. The logarithmic terms $a_{mn}^{{ F}}$ for $n\geq 1$ then can be expressed as
\begin{align}\label{eq:aFij}
a^F_{mn} = \frac{1}{j_{F}^n} \Bigl(
a_{mn}^F[\beta] + a_{mn}^F[\beta,\gamma_F] + a_{mn}^{F}[\beta,\Gamma_{\!F}]
\Bigr)\, ,
\end{align}
where the three terms result from the running of the coupling, the non-cusp and cusp pieces of the anomalous dimension of the given function.
The coefficients in \eq{aFij} can be obtained via the following recursion relations:
\begin{align}\label{eq:genLogs}
&a_{mn}^{F}[\beta] =
\frac{2}{n}\!\sum_{i = n}^{m-1}
i\, a^{F}_{i(n-1)}[\beta]\,
\beta^{(n_f)}_{m-i-1}\, ,
\;\;\,\quad 1\leq n\leq m -1\,, \nn
\\
&a_{mn}^{F}[\beta,\gamma_{F}] =
\frac{2}{n}\!\!\sum_{i = n-1}^{m-1}\!\!\!
i\, a_{i(n-1)}^{F}\![\beta,\gamma_{F}]
\beta^{(n_f)}_{m-i-1}\, ,
\;\; 2\leq n\leq m \,,
\nn\\
&a_{mn}^{F}[\beta,\Gamma_{\!F}] = \\
&\quad\quad
\frac{2}{n}\!\sum_{i = n-2}^{m-1}\!\!
i\, a_{i(n-1)}^{\!F}[\beta,\Gamma_{\!F}]\,\beta^{(n_f)}_{m-i-1}\, , \qquad
3\leq n\leq m+1
\, ,
\nn
\end{align}
with $m > 1$ in order to have a sensible upper limit.
The starting values of the three series (with $m \geq 1$) are given by
\begin{align}\label{eq:am012}
a^{ F}_{m0}[\beta] &= a^{ F}_{m0} \, , \qquad\quad
a^{ F}_{m1}[\beta,\gamma_{ F}] = \gamma^{ F}_{m-1} \, , \\
a^{ F}_{m2}[\beta,\Gamma_{\! F}] &=\frac{j_{{F}} }{2}\, \Gamma^{ F}_{\! m-1} \, ,
\nn
\end{align}
with $a^{ F}_{m0}[\beta,\gamma_{ F}]=a^{ F}_{m0}[\beta,\Gamma_{\! F}]=a^{ F}_{m1}[\beta,\Gamma_{\! F}]=0$.
Here, the integer $j_{ F}$ corresponds to the dimension of the momentum-space variable $q$ as it appears naturally in
logarithms,\footnote{Even though from \eq{RGEall} one can see that $j_{ F} = 1$ for all the functions we consider in our analysis, for sake of generality we have left it explicit in the formulae above.} as shown in \eq{posRGE2}. The dependence on $j_{ F}$ is factorized as in \eq{aFij} and only enters the $a^{ F}_{mn}[\beta,\Gamma_{\! F}]$ coefficients through the boundary condition in \eq{am012}. For the factorization functions in \eq{posRGE2} this is simply set to 1.
The constant terms of the SCET matrix elements up to NNLO are as follows:
\begin{align}\label{eq:FixedOrderCoeffs}
\{a_{i,0}^{H_Q} \}_{1\leq i\leq 2} &= \{9.372, \, 305.454\} \, , \nn \\
\{a_{i,0}^{J_{\!B,\tau_2}} \}_{1\leq i\leq 2} &= \{15.053, \, 310.954\} \, , \\
\{a_{i,0}^{S_{\tau_2}}\}_{1\leq i\leq 2} &= \{-13.159, \, -81.361\} \, . \nn
\end{align}

\section{MSR mass}
\label{app:msr}
The defining series for the standard $\MS$ mass $\overline m_t^{(6)}\equiv\overline m_t^{(6)}(\overline m_t^{(6)})$
reads
\begin{align}\label{eq:poleMSbardef}
\delta \mbar &\equiv m_t^{\rm pole} - \mbar_t^{(6)} \\
&= \mbar_t^{(6)}\,\sum_{i=1} \biggl[\frac{\alpha_s^{(6)}(\mbar_t^{(6)} )}{4\pi}\biggr]^i\, a_i^{(6)}(5,1) \, ,
\nn
\end{align}
where the notation $a_i^{(n_f)}(n_\ell, n_h)$ refers to the coefficient for $n_\ell$ massless and $n_h$ heavy flavors
with the running coupling expressed in the $n_f$-flavor scheme (note that in general $n_f\neq n_\ell + n_h$).
The defining relation for the MSR mass\footnote{We remind the reader that we adopt the `natural' MSR scheme as defined in Ref.~\cite{Hoang:2017suc} and refer to is as just the MSR mass.} is given by
\begin{align}\label{eq:poleMSRndef}
\delta m(R) &= m_t^{\rm pole} - m_t^{{\rm MSR},(5)}(R) \\
&= R\,\sum_{i=1} \biggl[\frac{\alpha_s^{(5)}(R)}{4\pi}\biggr]^i\, a_i^{(5)}(5,0) \, ,
\nn
\end{align}
where $n_h = 0$ signifies that the virtual self-energy corrections coming from top quark virtual loops have been integrated out.
Using these results we can write down the matching relation between the $\MS$ and the MSR masses at the scale $\mbar_t^{(6)}$:
\begin{align}
\label{eq:MSRnMSbarmatch}
m_t^{{\rm MSR},(5)}(\mbar_t^{(6)}) - \mbar^{(6)}_t &= \mbar^{(6)}_t\,
\sum_{i=1} \biggl[\frac{\alpha_s^{(5)}(\mbar^{(6)}_t)}{4\pi}\biggr]^i\, \Delta a_i^{(5)}\,,
\end{align}
with
\begin{align}\label{eq:DeltaAn}
\Delta a_i^{(5)} &= a_i^{(5)}(5,1) - a_i^{(5)}(5,0) \, .
\end{align}
The coefficients have following numerical values~\cite{Tarrach:1980up, Gray:1990yh,Chetyrkin:1999ys,Chetyrkin:1999qi,Melnikov:2000qh,Marquard:2007uj}:
\begin{align}
\label{eq:anAll}
&\{a_i^{(5)}(5,0) \}_{1\leq i\leq 3}
= \{5.333, \, 130.128, \, 4582.535
\} \, ,
\\
&\{a_i^{(5)}(5,1) \}_{1\leq i\leq 3}
= \{5.333, \, 131.785, \, 4699.703
\} \, .\nn
\end{align}

\section{Soft Gap subtraction schemes}
\label{app:gap}

We can generically start the construction of the gap subtraction series $\bar\delta$ that shall cancel the soft function renormalon in \eq{Spositionspace} by considering the following general condition:
\begin{align}
\label{eq:gapdef0}
&\frac{\df^n}{\df \log^n (iy)} \log \Bigl[ \tilde S_{\tau_2}^{(5)}(y,\mu_\delta)
e^{-2iy\bar{\delta}(\mu_\delta, R_s ;\, n, \xi) }
\Bigr]_{y =\frac{\xi}{iR_s}} =0
\,,
\end{align}
where $n \geq 0$ and $\xi \sim {\cal O}(1)$ is an auxiliary parameter.
This condition specifies a physical renormalon-free ``momentum-subtraction-like'' scheme which defines the soft function by imposing a condition on it at a point in position space.
Here $ \tilde S_{\tau_2}^{(5)}(y,\mu_\delta)$ is the $\MS$ soft function and $\mu_\delta$ is a reference renormalization scale that can be chosen independently of $R_s$.
Solving \eq{gapdef0} for $\bar\delta$ gives
\begin{align}\label{eq:gapdef01}
&\bar{\delta}(\mu_\delta, R_s \, ; n, \xi) \equiv
\frac{R_s}{2\,\xi}
\frac{\df^n}{\df \log (iy )^n} \log\Bigl[ \tilde S_{\tau_2}^{(5)}(y, \mu_\delta)
\Bigr]_{iy =\frac{\xi}{R_s}}
\, .
\end{align}
This defines a range of gap subtraction schemes for different choices of $n$ and $\xi$.
The renormalon in the soft function is not influenced by the terms depending on the cusp or non-cusp anomalous dimensions, and hence appears only in the terms $a^{S_{\tau_2}}_{ij}[\beta]$'s using the notation of \eq{aFij}.
For a given $n$, these terms enter the series for $\bar\delta$ at ${\cal O}(\alpha_s^{n+1})$. The choices $n=0, 1$ yield the two subtractions schemes given in \eqs{gapdef}{gapdef1}. We do not consider schemes with $n\ge 2$ where the $a_{ij}[\beta]$ enter at ${\cal O}(\alpha_s^3)$ and beyond.

The scheme used in Refs.~\cite{Hoang:2008fs,Abbate:2010xh} corresponds to the choice $\xi = e^{\gamma_E}$ and $\mu_\delta = \mu_S$, yielding \eq{gapdef1}.
Here we instead employ the $n = 0$ scheme in \eq{gapdef}, setting $\xi=1$ and $\mu_\delta = R_s$ instead, which makes the gap parameter $\overline \Delta$ independent of the renormalization scale $\mu_S$ of the soft function. This yields
\begin{align}\label{eq:deltaRsXi}
\bar{\delta}(R_s ) &\equiv
\bar{\delta}(R_s, R_s \, ; 0, 1)
=R_s
\sum_{i = 1} \biggl[\frac{\alpha^{(5)}_s(R_s)}{4\pi}\biggr]^i\,
d_{i0}
\,,\\
& \text{where } d_{i0} = \frac{1}{2}\sum_{j = 0}^{i+1} s_{ij} \gamma_E^j
\, ,\nn
\end{align}
where the $s_{ij}$'s are simply the constant terms and coefficients of powers of logarithms $\tilde L_S^j = \log^j(ie^{\gamma_E}y \mu)$ in the fixed-order expansion of $\log\bigl[S_{\tau_2}^{(5)}(y,\mu)\bigr]$ at ${\cal O}(\alpha_s^i)$ in \eq{Soft} (see \eq{lnSBcoeffs}).
We remind the reader that when $\bar\delta(R_s)$ is used in the factorization theorem, it is crucial that $\bar\delta$ is treated as a series expansion in $\alpha^{(5)}_s(\mu_S)$, the same coupling used in the series for the soft function. Therefore the final expression used for our analysis is
\begin{align}
\bar \delta (R_s)
&= R_s \sum_{i = 1}
\biggl[\frac{\alpha^{(5)}_s(\mu_S)}{4\pi}\biggr]^i\,
d_{ij} \log^j \Bigl(\frac{\mu_S}{R_s}\Bigr)
\, ,
\\
& \text{with } d_{ij}
=
\frac{2}{j}\sum_{k = j}^{i-1}
k\,d_{k(j-1)}\,
\beta^{(5)}_{i-k-1}
\,. \nn
\end{align}
Finally, we can relate the leading power correction $\Omega_1(R_s)$ between the two subtraction schemes:
\begin{align}\label{eq:Omega1OldNewDiff}
&\Omega_1(R_s)- \Omega_1^{\tiny \text{\,Refs.\cite{Hoang:2008fs,Abbate:2010xh}}}(R_s,R_s) \\
&\qquad = R_s \,[\,\bar \delta^{\tiny \text{Refs.\cite{Hoang:2008fs,Abbate:2010xh}}} (R_s, R_s) - \bar \delta (R_s)\,]
\nn\\
& \qquad= R_s \biggl\{8.357\frac{\alpha_s^{(5)}(R_s)}{4\pi} + 28.489\biggl[\frac{\alpha_s^{(5)}(R_s)}{4\pi}\biggr]^2 \biggr\}
\,. \nn
\end{align}

\section{R-evolution}
\label{app:Revolution}
We consider a generic perturbative series depending linearly and logarithmically on the scale $R$ that has the following form:\footnote{In this section it is assumed that $\alpha_s$ runs with $n_f$ active flavors.}
\begin{align}
\label{eq:fRdef}
f(R) = R \sum_{i = 1}^\infty \biggl[\frac{\alpha_s(R)}{4\pi}\biggr]^i\, f_i \, .
\end{align}
The evolution equation of $f(R)$ with respect to $R$ is then given by
\begin{align}
\label{eq:fRRGE}
\frac{\df f(R)}{\df \ln R } =R\frac{\df f(R)}{\df R } =R\sum_{n = 0}^\infty \gamma_n^{f,R}
\biggl[\frac{\alpha_s(R)}{4\pi}\biggr]^{n+1} \, ,
\end{align}
with the $R$ anomalous dimension coefficients being
\begin{align}\label{eq:gammaRn}
\gamma_n^{f,R}
&= f_{n+1} -
2 \sum_{j = 0}^{n-1} (n-j)\beta_j f_{n-j} \,,\, \quad (n \geq 1)\,,
\end{align}
where $\gamma_0^{f,R} = f_1$. The crucial aspect of using the RG equation in \eq{fRRGE} is that the $R$-evolution anomalous dimension
in Eq.~(\ref{eq:gammaRn}) is ${\cal O}(\Lambda_{\rm QCD})$ renormalon-free, if the series in \eq{fRdef} contains
an ${\cal O}(\Lambda_{\rm QCD})$ renormalon~\cite{Hoang:2008yj,Hoang:2017suc}. This is a fundamental ingredient in the construction of renormalon subtractions in distributions where the subtraction scale depends on the value of kinematic quantities~\cite{Hoang:2007vb,Hoang:2009vu}.
The solution of the $R$-evolution equation is straightforward and given by
\begin{align}\label{eq:Revol}
f(R_1)\! -\! f(R_0)
= \sum_{n = 0}^\infty
\gamma_n^{f,R} \!\!
\int_{R_0}^{R_1}\!\! \df R
\biggl[\frac{\alpha_s(R)}{4\pi}\biggr]^{n+1}
\! .
\end{align}
Using \eq{gammaRn} we can derive the anomalous dimensions for $R$-evolution of the MSR mass and the gap subtractions by using the following values for $f_i$:
\begin{align}
\text{$\delta m(R)$, \eq{poleMSRndef}}&: \qquad f_i = a_i^{(5)}(5,0) \, , \\
\text{$\bar \delta (R_s)$, \eq{deltaRsXi}}&: \qquad f_i = d_{i0}\, . \nn
\end{align}
Note that the $R$-evolution equations for the MSR mass $m_t^{\rm MSR}$ and the gap parameter $\overline \Delta(R_s)$ in \eqs{RRGE}{RsRGE} are defined with an additional minus sign of the evolution equations of the corresponding subtraction terms which are referred to in \eq{Revol}. For the $R$-anomalous dimensions we get
\begin{align}\label{eq:gammaR}
\{\gamma_i^{\rm MSR}\}_{0\leq i\leq 3} &= \{5.333, 48.350, 179.501\}\,,\\
\{\gamma_i^{\Delta}\}_{0\leq i\leq 2} &= \{-8.357, 55.693\}\,.\nn
\end{align}
\section{Nonperturbative model function}
\label{app:model}
We use the nonperturbative model function $F(k)$
that has the form
\begin{align} \label{eq:basis1}
F(k) = S_\tau^{\rm mod}(k,\lambda,\{c_i\}) \equiv
\frac{1}{\lambda} \Biggl[\, \sum_{n=0}^N
c_n \, f_n\biggl(\frac{k}{\lambda}\biggr)
\Biggr]^2,
\end{align}
where the basis functions are~\cite{Ligeti:2008ac}
\begin{align} \label{eq:basis2}
f_n(z) & = 8 \sqrt{\frac{2 z^3 (2 n + 1)}{3}}\,\, e^{-2 z}\, P_n\bigl(g(z)\bigr),
\\
g(z) & = \frac{2}{3}\bigl[ 3 - e^{-4 z}\,(3 + 12 z + 24 z^2 + 32 z^3)\bigr] -1\,,\nn
\end{align}
and $P_n$ are Legendre polynomials. For $\sum_i c_i^2=1$ the norm of $S^{\rm
mod}_\tau(k)$ is unity, i.e.\ $\Omega_0=1$. The choice of basis in
\eqs{basis1}{basis2} depends on specifying one dimensionful parameter $\lambda$ which is characteristic for the width of the soft function.
Following Ref.~\cite{Abbate:2010xh}, in our analysis we set \mbox{$N=2$} and $c_1=0$,
such that the first moment using \eq{O1bar} is given by
\begin{align} \label{eq:Omega12}
\overline \Omega_1 &= \Delta +
\frac{\lambda}{2} \bigl[ c_0^2 + 0.201 c_0 c_2 + 1.100 c_2^2 \bigr]\,,
\end{align}
where the normalization condition $c_0^2 + c_2^2 =1$ can be used to express $c_0>0$ in terms of $c_1$.
For our numerical analyses we take \mbox{$c_2 = 0.05$}, $\Delta = 0.1$ GeV, and the value of $\overline \Omega_1$ then fixes $\lambda$.
The $\Omega_1(R_s)$ including gap-running can be evaluated using \eqs{Omega1Rs}{deltaRsXi}. For a reference scale of $R_s = 2$\,GeV we find:
\begin{align}\label{eq:Omega1Rs2GeV}
\Omega_1 (
2\,\mbox{\rm GeV} )
&=
\overline \Omega_1 -
(2 \,\mbox{\rm GeV})
\sum_{i = 1}^{2} \biggl[\frac{\alpha_s^{(5)}(2\,\mbox{\rm GeV})}{4\pi}\biggr]^i\,
\delta_{i0}
\\
&= \overline \Omega_1 + 0.463\,\mbox{\rm GeV} \, .\nn
\end{align}

\bibliographystyle{apsrev4-1}

\bibliography{../top3}

\newpage
\clearpage

\end{document}